\documentclass{emulateapj}

\newcommand{\fig}[1]{Figure~\ref{#1}}
\newcommand{\tab}[1]{Table~\ref{#1}}

\newcommand{\ib}{$i'$}
\newcommand{\zb}{$z'$}

\newcommand{\sol}{$_{\odot}$}

\newcommand{\lya}{Ly$\alpha$}
\newcommand{\ha}{H$\alpha$}
\newcommand{\dust}{A$_{1200}$}
\newcommand{\SII}{[S\,{\sc ii}]}

\def\arcs{\hbox{$^{\prime\prime}$}}
\def\arcm{\hbox{$^{\prime}$}}

\shorttitle{\lya Galaxies at z $\sim$ 4.5}
\shortauthors{Finkelstein et al.}

\begin{document}
\title{Lyman Alpha Galaxies: Primitive, Dusty or Evolved?\altaffilmark{1}}

\author{Steven   L.    Finkelstein\altaffilmark{2}$^{,}$\altaffilmark{3},   James   E.    Rhoads,   Sangeeta
Malhotra\altaffilmark{4} \& Norman Grogin\altaffilmark{4}$^{,}$\altaffilmark{5}}   
\altaffiltext{1}{Based in part on observations taken at the Cerro Tololo Inter-American Observatory.}
\altaffiltext{2}{Department  of  Physics,  Arizona  State University,  Tempe, AZ  85287;  stevenf@asu.edu} 
\altaffiltext{3}{Department of Physics, Texas A\&M University, College Station, TX 77843}
\altaffiltext{4}{School of Earth and Space Exploration,  Arizona  State University,  Tempe, AZ  85287} 
\altaffiltext{5}{Space  Telescope Science Institute, 3700  San Martin Drive, Baltimore,  MD 21218}

\begin{abstract}
We present stellar population modeling results for 10 newly discovered Lyman alpha emitting galaxies (LAEs), as well as four previously known LAEs at z $\sim$ 4.5 in the Chandra Deep Field -- South.  We fit stellar population models to these objects in order to learn specifically if there exists more than one class of LAE.  Past observational and theoretical evidence has shown that while many LAEs appear to be young, they may be much older, with \lya~EWs enhanced due to resonant scattering of \lya~photons in a clumpy interstellar medium (ISM).  Our results show a large range of stellar population age (3 -- 500 Myr), stellar mass (1.6 $\times$ 10$^{8}$ -- 5.0 $\times$ 10$^{10}$ $M$\sol) and dust extinction (\dust~= 0.3 -- 4.5 mag), broadly consistent with previous studies.  With such a large number of individually analyzed objects, we have looked at the distribution of stellar population ages in LAEs for the first time, and we find a very interesting bimodality, in that our objects are either very young ($<$ 15 Myr) or old ($>$ 450 Myr).  This bimodality may be caused by dust, and it could explain the \lya~duty cycle which has been proposed in the literature.  We find that eight of the young objects are best fit with a clumpy ISM.  We find that dust geometry appears to play a large role in shaping the SEDs that we observe, and that it may be a major factor in the observed \lya~equivalent width distribution in high redshift \lya~galaxies, although other factors (i.e. outflows) may be in play.  We conclude that 12 out of our 14 LAEs are dusty star-forming galaxies, with the other two LAEs being evolved galaxies.
\end{abstract}


\section{Introduction}
While high-redshift galaxies can be hard to observe due to dimming with distance, narrowband selection of \lya~galaxies has proven a very efficient method to select high-redshift galaxies based on a strong emission line (e.g., Rhoads  et al  2000, 2004;  Rhoads \&  Malhotra 2001; Malhotra \& Rhoads 2002; Cowie \&  Hu 1998; Hu et al 1998, 2002, 2004; Kudritzki et al 2000; Fynbo, Moller, \& Thomsen 2001; Pentericci et al 2000; Ouchi et al 2001, 2003, 2004; Fujita et al 2003; Shimasaku et al 2003, 2006;  Kodaira et  al 2003;  Ajiki et al  2004; Taniguchi  et al 2005; Venemans et al 2002, 2004; Gawiser et al. 2006; Lai et al. 2007, 2008; Nilsson et al. 2007; Finkelstein et al. 2008a).  These objects are of interest, as over 40 years ago Partridge \& Peebles (1967) proposed that they may be signs of primitive galaxies in formation.  This was easy to understand, as \lya~photons are copiously produced in star formation regions, and we would expect the first galaxies to be undergoing periods of extreme star formation.  However, not until recently have we had the data to verify this assumption.

The availability of stellar population modeling codes has allowed the derivation of physical parameters of galaxies from photometry alone.  In the past few years (thanks to large surveys such as the Great Observatories Origins Deep Survey (GOODS)), broadband photometry of LAEs has become deep enough to compare objects to these models, learning about such parameters as stellar population age, stellar mass, and dust extinction.  First results from these studies were unsurprising, as stacking analyses showed that an average LAE was young ($\sim$ 10 -- 100 Myr), low mass (10$^{7 - 8}$ $M$\sol) and dust free (e.g., Gawiser et al. 2006; Finkelstein et al. 2007; Lai et al. 2008).  Recently, deeper data has allowed the comparison of individual high-redshift LAEs to models for the first time, as chronicled in Chary et al. (2005), Pirzkal et al. (2007), Lai et al. (2007) and Finkelstein et al. (2008a).  These studies have shown a wide range of results, with LAE ages from 1 -- 1000 Myr, masses from 10$^{6 - 10}$ $M$\sol~and, most surprisingly, dust extinction with A$_{V}$ up to 1.3 mag.  These results show that while some LAEs may be young and dust-free, many are dusty, and some are even evolved (i.e. old and high-mass).

This raises an interesting question, as how can an evolved stellar population produce enough \lya~photons to be picked up by a narrowband selected survey?  Many scenarios have been proposed (i.e. zero metallicity, top-heavy initial mass function etc.), but these are rather extreme, and given the amount of dust extinction we see, they would be unlikely to produce enough \lya~photons to explain the observed excesses.  We have thus decided to observationally test a scenario developed theoretically by Neufeld (1991) and Hansen \& Oh (2006), where the \lya~equivalent width (EW) actually gets {\it enhanced} due to a dusty interstellar medium (ISM).  This is counter-intuitive, as dust will strongly suppress any ultraviolet (UV) photons.  The news is even worse for \lya~photons, as they resonantly scatter off of neutral hydrogen atoms, thus their mean-free-paths are rather small, vastly increasing their chances of encountering a dust grain.  However, as Neufeld, Hansen \& Oh suggest, an ISM that is very clumpy could actually prevent the \lya~photons from seeing the dust at all.  This can happen if the dust and neutral hydrogen are evenly mixed together in clumps, with a tenuous ionized medium separating the clumps.  In this geometry, the \lya~photons stand a very high chance of being resonantly scattered at the surface of these clumps, spending most of their time in the inter-clump medium.  In this manner, the \lya~photons are effectively screened from ever encountering a dust grain, as the gas and dust are in the same geometry, so the \lya~gets shielded from encountering dust.  The story is different for continuum photons, as they are not resonantly scattered, thus they will penetrate deeply into a clump, with a strong chance of being scattered or absorbed.  As EW is a measure of line-to-continuum flux, this effectively enhances the observed EW over that due to the stars (or more accurately, the stars interaction with the interstellar hydrogen in their immediate surroundings).  Note that the \lya~flux itself is not being increased, rather the \lya-to-continuum ratio is being enhanced (see Finkelstein et al. 2008a, \S 3.2 for a detailed explanation of this scenario).

In Finkelstein et al. (2008a; hereafter F08a), we analyzed a sample of four LAEs, fitting model spectra to their SEDs, searching for proof that this type of ISM exists.  Three of our objects were best fit by young (5 Myr) dusty (\dust~$\sim$ 1 -- 2 mag) stellar populations, similar to those seen by Pirzkal et al. (2007) at z $\sim$ 5 and Chary et al. (2005) at z $\sim$ 6.5.  However, the fourth object was best-fit by an old (800 Myr) stellar population, with 0.4 mag of dust arrayed in a clumpy ISM, showing evidence of dust enhancement of the \lya~EW.  Also interesting was the age distribution, as these galaxies were either very young or very old, although this distribution was hard to quantify with such a small sample.  Using our larger sample, we will now see if the absence of ``teenage'' LAEs is real, and see if most LAEs are primitive, dusty or evolved.  Further detection of this dust-enhancement of the \lya~EW can help explain the larger than expected EWs seen in many LAEs (e.g., Kudritzki et al. 2000; Malhotra \& Rhoads 2002; Finkelstein et al. 2007).

This paper is organized as follows.  In \S 2 we present our observations, including our object selection and redshift information where applicable.  In \S 3 we present our stellar population models.  Our best-fit results for each object are presented in \S 4, and we discuss the implications of these results in \S 5, including suggestions for future improvement.  We present our conclusions in \S 6.  In this paper we assume Benchmark Model cosmology, where $\Omega_{m}$ = 0.3, $\Omega_{\Lambda}$ = 0.7 and H$_{0}$ = 0.7 (c.f. Spergel et al. 2007).  All magnitudes in this paper are listed in AB magnitudes (Oke \& Gunn 1983).

\section{Data Handling}

\subsection{Observations}
In F08a, we published a study based on narrowband imaging of the GOODS Chandra Deep Field -- South (CDF--S; RA 03:31:54.02, Dec -27:48:31.5  (J2000)) obtained at the Blanco 4m telescope at Cerro Tololo InterAmerican Observatory (CTIO) with the MOSAIC II camera (MOSAIC II is a mosaic CCD camera, with eight 2k$\times$4k CCDs spanning 36\arcm$\times$36\arcm on the sky).  These data were taken in the NB656 (\ha) filter, and resulted in the discovery of four LAEs at z = 4.4.  In semester 2007B, we applied for, and were awarded, three nights at Blanco with MOSAIC II (two whole nights and two $\frac{1}{2}$ nights) to obtain observations in two 80 \AA~wide narrowband filters adjacent to NB656.  We obtained five hours of observations in the NB665 (\ha+80) filter, and 5.5 hours in the NB673 (\SII) filter, which can be used to select LAEs at z = 4.47 and z = 4.52 respectively.  Conditions were photometric over the whole run, with a typical seeing of 0.9\arcs.

The CTIO data were reduced with IRAF\footnote[1]{IRAF is distributed by the National Optical Astronomy Observatory (NOAO), which is operated by the Association of Universities for Research in Astronomy, Inc.\ (AURA) under cooperative agreement with the National Science Foundation.} \citep{tody86,tody93}, using the MSCRED \citep{valtod,val98} reduction package, following the method set forth in Rhoads et al. (2000, 2004).  First, we performed the standard image reduction steps of overscan subtraction and bias subtraction, followed by flat--fielding using dome flats obtained during our run.  Cross-talk was also removed between chip pairs sharing readout electronics.  We derived a supersky flat from the science data, and used this to remove the residual large--scale imperfections in the sky.  The world coordinate systems (WCS) of individual frames were adjusted by comparing the frames to the astrometry from the USNO--B1.0 catalog.  Cosmic rays were rejected using the algorithm of Rhoads (2000) and satellite trails were manually flagged and excluded from the final stacked image.  The final stack for each filter was made using mscstack, using scaling from mscimatch.  See Rhoads et al. (2000) and Wang et al. (2005) for further details on the data reduction process.  The seeing and zeropoint of the final stack was 0.92\arcs (0.91\arcs) and 29.811 (29.847) for the NB665 (NB673) filter.

\subsection{Object Extraction}
To select LAEs in our field, we require broadband photometry encompassing the \lya~line to calculate a narrowband excess, as well as a broadband filter blueward of the line verifying that the flux is indeed extinguished due to intervening inter-galactic medium (IGM) material.  We used $B$ and $R$ broadband data from the ESO Imaging Survey (EIS; Arnouts et al. 2001), which obtained deep U'UBVRI data of 0.25 deg$^{2}$ surrounding the CDF--S.  The 5 $\sigma$ depth in our bands of interest are B$_{AB}$ = 26.4 and R$_{AB}$ = 25.5.  While there is deep, high-resolution {\it Hubble Space Telescope} ({\it HST}) data in this field, in order to do our selection we would have to smooth these data to the resolution of our ground-based data\footnote[2]{If we run our selection with the full resolution {\it HST} data, we run into a problem with crowded sources, where two objects in the {\it HST} image are not resolved separately in the narrowband images.  This results in a computed narrowband excess, where one may not exist.}, and at that point, the EIS data is marginally deeper, and it has the benefit that we can select LAEs over a larger area.  The EIS $B$ and $R$-band images were registered separately to both narrowband images using the IRAF tasks wcsmap and geotran.

We extracted objects from our stacked images using the SExtractor software (Bertin \& Arnouts 1996) in two-image mode.  In this mode, the software first detects objects in the narrowband (detection) image, and then extracts the flux from those positions in the second image.  Using both the NB665 and NB673 as the detection image, we extracted fluxes from the EIS $B$ and $R$-bands, as well as all three narrowbands (including the NB656 image from F08a).  In order to reduce the number of false positive detections, we first extracted objects using a set of SExtractor parameters used in the Large Area Lyman Alpha (LALA; Rhoads et al. 2000) Survey.  We then selected LAE candidates (see \S 2.3) using these catalogs.  We then iterated, changing the SExtractor detection parameters, reducing the number of objects detected while still detecting the previously identified candidates.  This iterative step resulted in a $\geq$ 33\% reduction in detected objects over the whole field, drastically reducing the number of spurious sources.  Our final detection parameters were DETECT$\_$MINAREA = 9 pixels ($\sim$ size of the seeing disk) and DETECT$\_$THRESH = 0.50 (NB665) and 0.95 (NB673).  We then used these parameters to create our final catalogs for narrowband selected sources in the narrowband, $B$ and $R$-band images.

\subsection{\lya~Galaxy Selection}
We use the selection criteria from Malhotra \& Rhoads (2002; hereafter MR02) to select \lya~galaxy candidates from our catalogs.  For an object to be considered as a candidate, we demand a 5 $\sigma$ significance detection in the narrowband, a 4 $\sigma$ significance narrowband flux excess over the $R$ band, a factor of 2 ratio of narrow-broad flux, and no more then 2 $\sigma$ significant flux in the $B$-band.  The first three criteria ensure that it is a significant narrowband detection with a significant narrowband excess, while the last criterion checks that it is at z $\sim$ 4 (sources at z $\gtrsim$ 3.5 will have their $B$-band flux reduced due to IGM absorption).  Application of these selection criteria in the LALA survey resulted in an AGN contamination fraction of $<$ 5\% (Wang et al. 2004), and a follow-up spectroscopic confirmation success rate of $>$ 70\% (Dawson et al. 2004).

\begin{figure}
\epsscale{1.2}
\plottwo{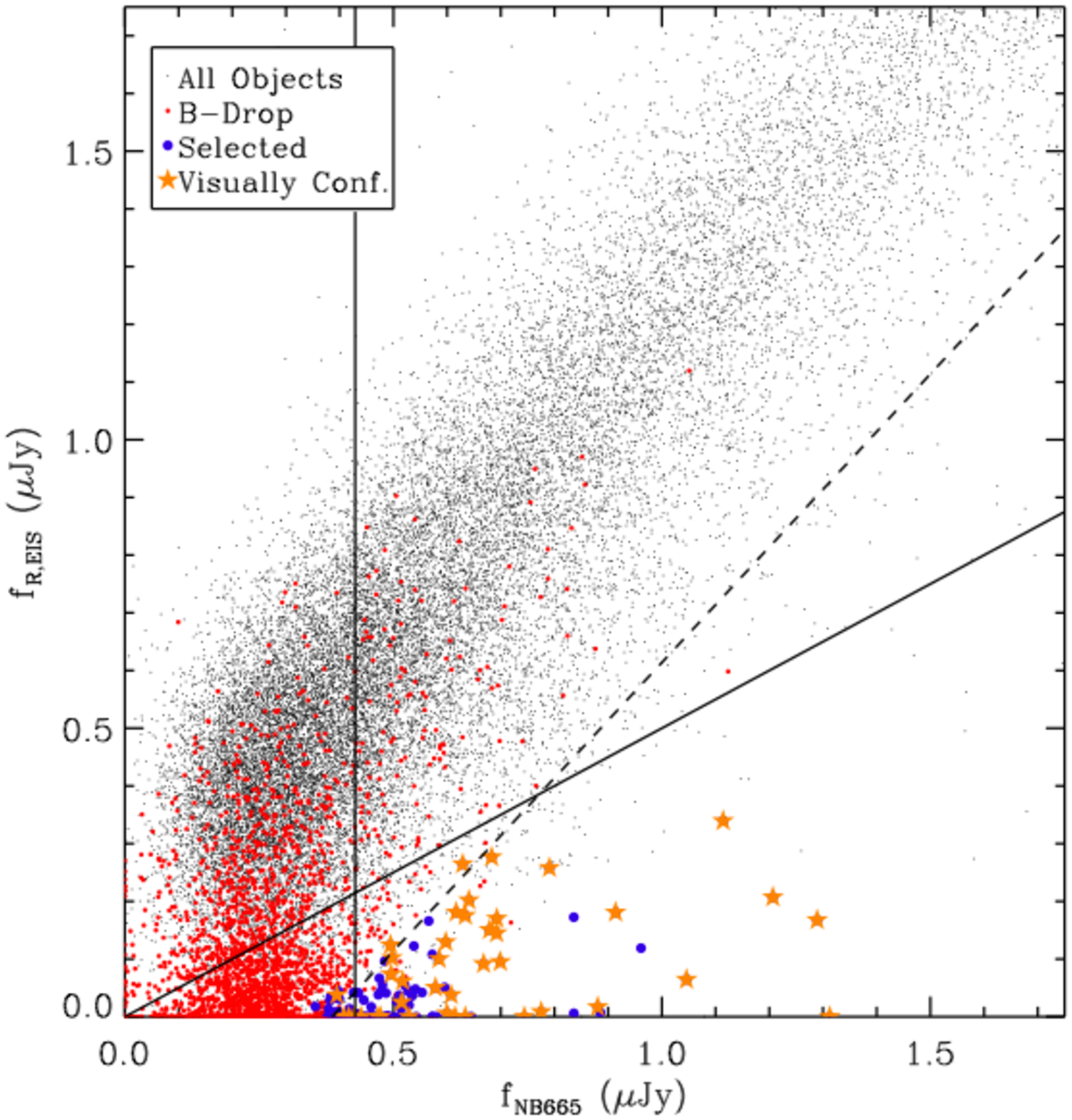}{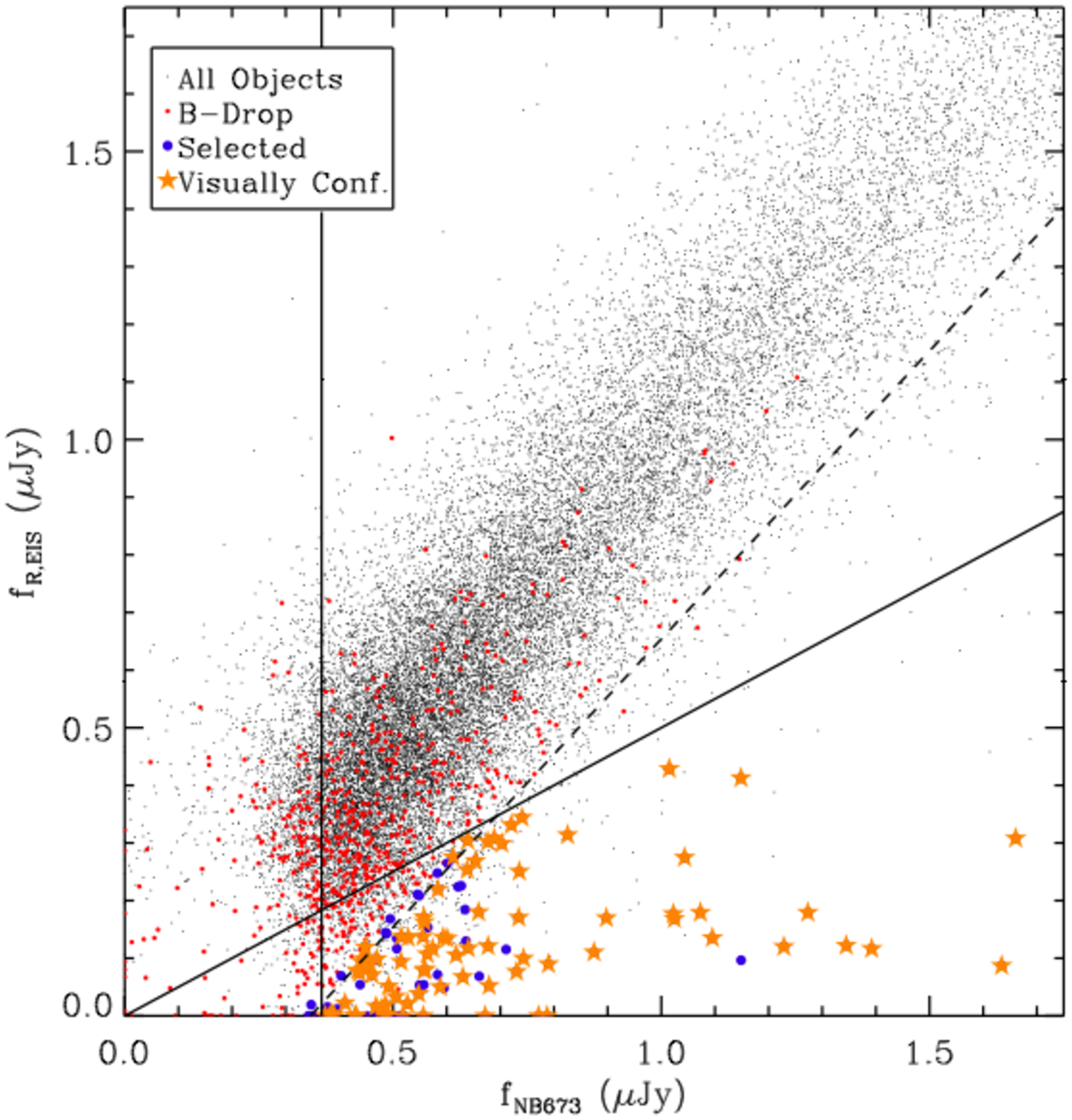}
\caption{Flux selection planes for objects detected in NB665 (left) and NB673 (right), with broadband (EIS $R$) flux on the vertical axis and narrowband flux on the horizontal axis.  The three lines represent three of the selection criteria.  Any object to the right of the vertical line satisfies the 5 $\sigma$ narrowband detection requirement (computed using the mean narrowband flux error from all objects).  Objects below the solid diagonal line have a narrow-broad flux ratio of $>$ 2, and objects below the dashed diagonal line satisfy the 4 $\sigma$ narrowband excess criteria.  Black dots represent all objects extracted from the image, which red dots represent those that satisfy the fourth selection criteria, of a less than 2 $\sigma$ significant B-band flux.  Blue dots represent those objects which satisfy all four selection criteria, and gold stars are those candidates which were confirmed to be real upon visual inspection of the image.  While there should be no blue dots above the selection lines, and likewise no red dots below, there are a few because the lines were computed using the mean errors from the entire sample, while the selection criteria were applied to each object using their individual uncertainties.}\label{fluxselect}
\end{figure}

Selecting objects in the overlap region between our narrowband data and the EIS data, we find 130 LAE candidates in the NB665 image and 126 in the NB673 image.  While these objects all satisfied our selection criteria, it is possible that some of them could be noise spikes or other types of spurious sources (i.e. near the wings of bright stars).  As a last check, we visually inspected the narrowband images at the positions of each of these 256 candidates, ensuring that a real object (i.e. detectable above the noise by eye) existed at these positions.  Removing from our sample sources that did not appear to be real, we find a total of 42 good LAE candidates in the NB665 image, and 85 in NB673.  \fig{fluxselect}~shows our flux selection plane, highlighting the selection criteria and visually confirmed LAE candidates.

While the EIS data worked very well for our selection, we require the higher-resolution and broader wavelength coverage of the GOODS {\it HST} Advanced Camera for Surveys (ACS; Ford et al. 2000) and {\it Spitzer} Infrared Array Camera (IRAC; Fazio et al. 2004) data to be able to fit stellar population models to these objects.  For our analysis, we used the updated version of the GOODS CDF--S catalog (v1.9; Giavalisco et al. 2008 in prep) which has deeper observations in the \ib~and \zb-bands, resulting in lower error bars for those two bands.  The IRAC data do not yet have a public catalog, but a catalog has been created using the TFIT software package (Laidler et al. 2007). TFIT uses the spatial positions and morphologies of objects in an image with higher angular resolution (ACS z'-band in our case) to construct object templates, which are then fitted to a lower resolution image, solving for the object fluxes as free parameters. Using extensive experiments on both simulated and real data, Laidler et al. (2007) have shown that this template-fitting method measures accurate object photometry to the limiting sensitivity of the image.  Laidler et al. (2007) use this method to produce a GOOD ACS-IRAC multiband catalog, which we use for our Spitzer fluxes (N. Grogin, private communication).  One possible contaminant is that objects could show up in the IRAC data which were not visible in the ACS data.  These extremely red objects (EROs) have been studied extensively by Yan et al. (2004), and they find a number density of 4.55 $\times$ 10$^{-4}$ arcsec$^{-2}$ for EROs in the Hubble Ultra Deep Field.  Assuming that the region of possible contamination for a given LAE is equal to double the IRAC pixel size (a box 2.4\arcs on a side), we find a ``contamination'' area of 5.76 arcsec$^{2}$ per object.  We would thus expect to find 0.0026 EROs within the bounds of one of our objects, or a 0.26\% chance of contamination.  We thus do not believe that EROs are significantly affecting our results.

To find which of our LAE candidates resided in the smaller GOODS CDF--S area, we performed position matching between our candidates and the GOODS v1.9 catalog, using a matching radius of 0.6\arcs.  We found that 17 total LAE candidates were covered by the GOODS data, 5 selected in NB665 and 12 in NB673.  While each of these objects satisfied our $B$-band deficiency criteria, four of them had visible flux in the $B$-band image, thus these four were thrown out of the sample as low-redshift interlopers, leaving us with 13 total LAE candidates (2 in NB665 and 11 in NB673).

\subsection{Candidate Redshifts}
While the {\it HST} Probing Evolution and Reionization Spectroscopically (PEARS; Rhoads et al. 2008) survey has obtained grism spectra over much of the CDF--S, none of the seven candidate LAEs which were covered by PEARS yielded any definitive information from their spectra, which were significantly noise dominated.  We then searched through two public catalogs, MUSIC (Grazian et al. 2006) and FIREWORKS (Wuyts et al. 2008) for photometric (photo-z) and spectroscopic (spec-z) redshifts for our candidates.  Six of our candidates had photo-z's from either MUSIC or FIREWORKS (when an object had a photo-z in both surveys, we use the value from FIREWORKS, as they published the 1 $\sigma$ errors on each photo-z), and two of these six also had published spectroscopic redshifts.  

The first of these two had a spec-z of 0.508 from the VIMOS VLT Deep Survey (VVDS; Le Fevre et al. 2004).  This is an interesting redshift, as we would expect our two main interlopers to be [OII] and [OIII] emitters, at redshifts $\sim$ 0.8 and 0.3 respectively.  However, Le Fevre et al. give this redshift a quality value of 2, meaning that it only has a 75\% confidence level.  Furthermore, the photo-z from FIREWORKS is slightly higher, at 0.68 $\pm$ 0.07, thus we believe that this object is consistent with an [OII] emitting interloper, and we exclude it from further analysis.  

The second of these objects has a spec-z of 4.00 from the FORS2 spectroscopic catalog (v3.0; Vanzella et al. 2008).  This redshift was given a ``C'' quality, meaning that it is a ``potential redshift.''  We were able to download and look at the 1D spectrum, it does appear that there is a possible continuum break at $\sim$ 6200 \AA, with two emission lines redward of the break.  They identified the first line as \lya, putting it at z = 4.0.  The second of the emission lines is the one which fell in our narrow-band filter, and if this redshift is correct, then we detected CII emission ($\lambda$1335 \AA).  While this is possible, it is unlikely, thus we think that two scenarios could be going on.  This break could be due to the \lya~forest, and the first emission line is just a noise spike, meaning that we are indeed measuring \lya~emission at z $\sim$ 4.5 (this is tenuously verified by the photo-z of 4.16 $\pm$0.16, which is $\sim$ 2 $\sigma$ from 4.5).  Or, equally likely, this could be the 4000 \AA~continuum break, meaning that we are detecting Balmer line emission from H$\beta$ ($\lambda$4861 \AA), H$\gamma$ ($\lambda$4341 \AA), or H$\delta$ ($\lambda$4102 \AA) at z = 0.4 to 0.6.  Given these doubts, we decided to remove this object from the sample.

The four other candidate LAEs with redshifts had calculated photo-z's of 0.54, 3.82, 4.25 and 4.52.  Of these, we exclude only the first from further analysis given its low photo-z.  While its $B$-band flux does not appear upon visual inspection of the $B$-band image, it does appear quasi-well detected in a plot of its spectral energy distribution (SED), meaning that this object could be either an [OII] or [OIII] emitter.

\begin{figure}
\epsscale{2.1}
\begin{center}\plottwo{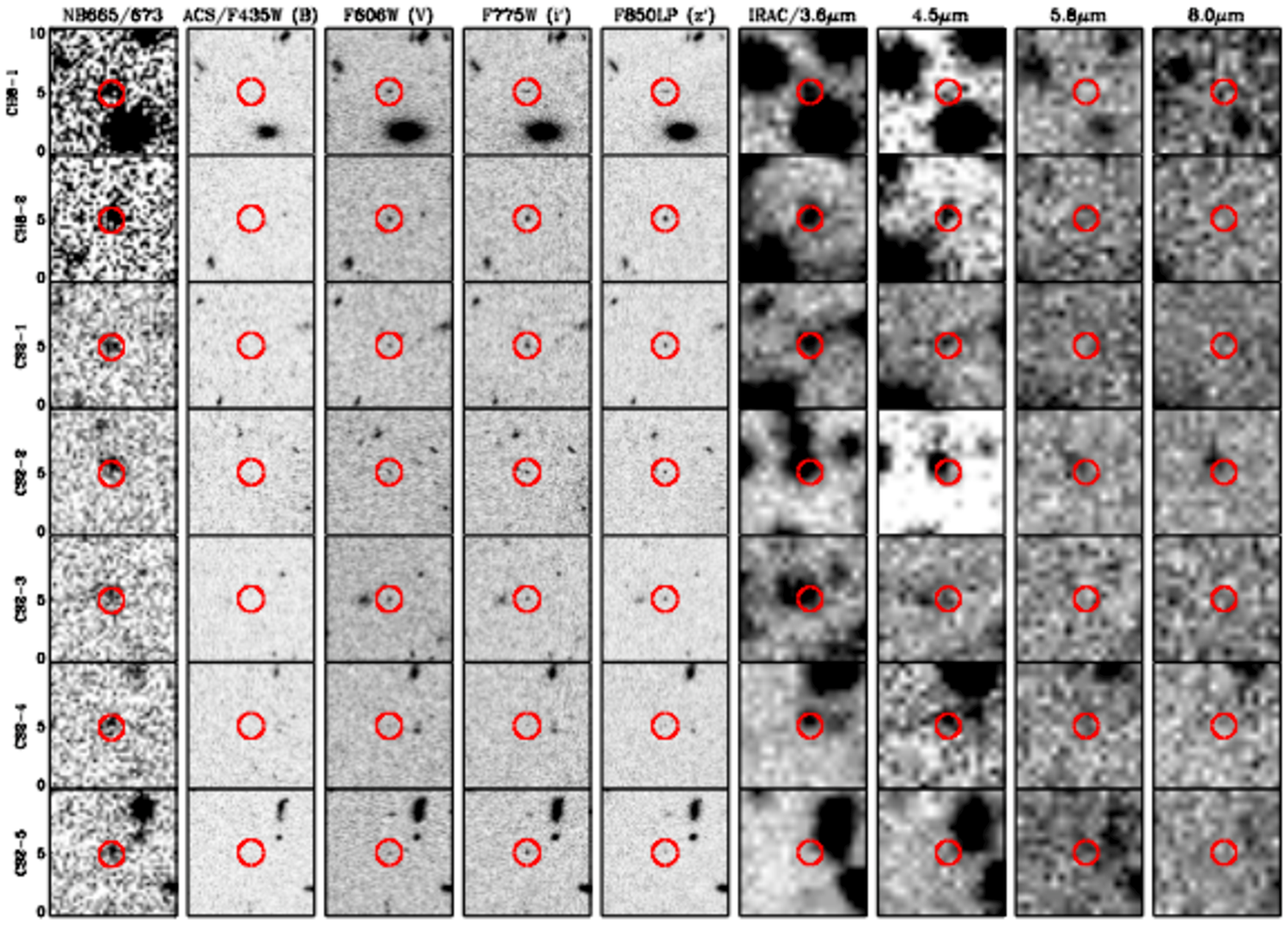}{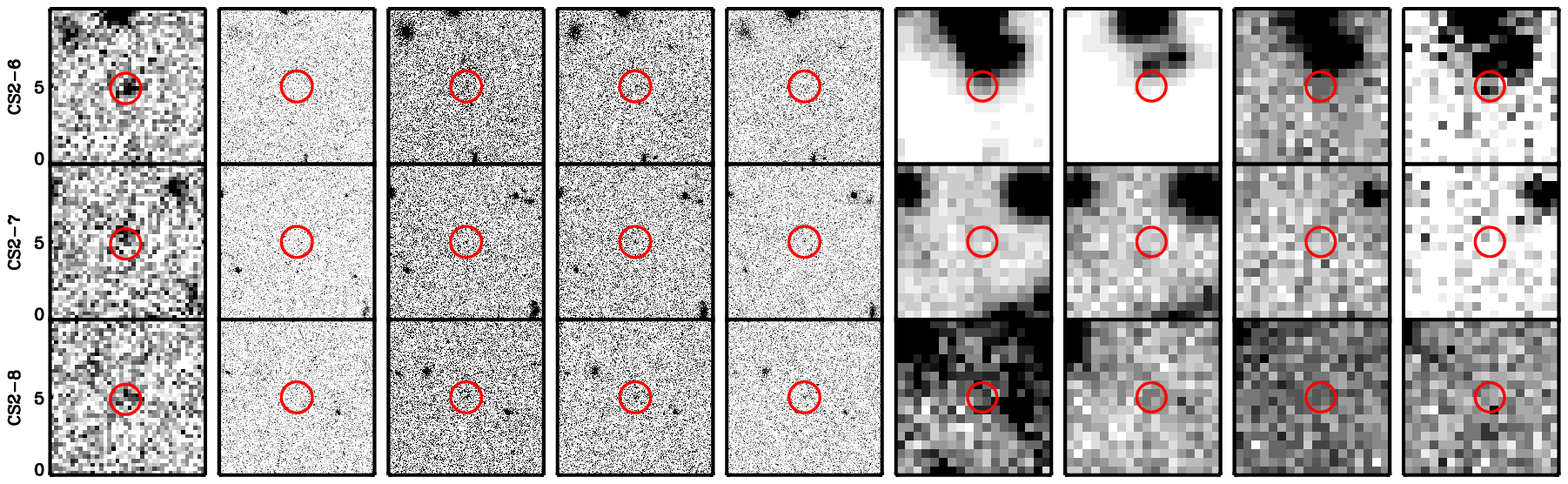}\end{center}
\caption{10\arcs~cutouts of our 10 new objects (see F08a for a similar figure for the original four objects).  The circles are 2'' in diameter, and are drawn to highlight the object.  The columns from left-to-right:  narrowband image (NB665 or NB673 depending on the object); {\it HST}/ACS $B$, $V$, $i'$ and $z'$; {\it Spitzer}/IRAC 3.6, 4.5, 5.8 and 8.0 $\mu$m.  All objects are easily visible in the narrowband detection images, and all look to have a brighter narrowband than V-band image by eye.  All objects are undetected in the IRAC 5.8 and 8.0 $\mu$m images.  The lack of detectable $B$-band flux illustrates the high-redshift nature of these objects.  While many objects are crowded in the IRAC 3.6 and 4.5 $\mu$m images, our use of the TFIT algorithm uses the ACS $z'$ high resolution image as a template to extract the fluxes of our crowded sources.}\label{stamps2}
\end{figure}

Lastly, while narrow-band surveys typically result in a low fraction of active galactic nuclei (AGN), they are a possible contaminant as they frequently show strong \lya~emission.  In order to check for AGN, we examined the {\it Chandra} 1 Ms X-ray catalog (Giaconni et al. 2001), performing position matching between our objects and this X-ray catalog.  We found no matches out to a search radius of 5\arcs, thus we believe that none of our final sample of 10 candidate LAEs contain AGNs.  \fig{stamps2}~shows cutout stamps of our 10 candidates at the relevant wavelengths.

\section{Stellar Population Modeling}
To learn about the physical properties of our objects, we computed model stellar population spectra using the software from Bruzual \& Charlot (2003; hereafter BC03).  We used a similar procedure as in F08a, computing model BC03 spectra from a grid of metallicities, star formation histories (SFHs), and stellar population ages.  We used five different metallicities, ranging from Z$_{absolute}$ = 0.0001 (0.005$Z$\sol) to 0.02 ($Z$\sol).  We used an exponentially decaying SFH, with the characteristic decay time, $\tau_{SFH}$, ranging from 10$^{5}$ - 4 $\times$ 10$^{9}$ yr.  The minimum value of $\tau_{SFH}$ is much shorter than our youngest possible model age, thus it is essentially a burst, or simple stellar population (SSP).  The maximum $\tau_{SFH}$ is much older than the age of the Universe at z = 4.4, thus it is simulating a continuous SFH.  The three intermediate values, 10$^{6}$, 10$^{7}$ and 10$^{8}$ yr represent true exponentially decaying SFHs.  We used a grid of 48 stellar population ages (t$_{pop}$), ranging from 1 Myr - 1.434 Gyr (the closest age point to the age of the Universe at z = 4.5), with 30 of the ages being $\leq$ 100 Myr.  In order to model the clumpy ISM scenario, we require our model spectra to contain \lya~emission lines.  While the BC03 code does not compute line emission, it does compute the number of ionizing photons at each age, and from this we can derive the \lya~emission line flux using Case B recombination.  We also included $H\alpha$ emission at a value of $\sim$ 11\% of the \lya~strength (Case B), as $H\alpha$ emission is typically strong in star forming galaxies, and at this redshift can dominate the 3.6 $\mu$m flux.

To apply dust attenuation to our spectra we used the dust law from Calzetti et al. (1994), which is applicable to starburst galaxies.  We attenuated our model spectra by 21 different amounts of dust extinction at 1200 \AA, from \dust~= 0.0 to 5.0.  While we allowed all continuum wavelengths to be attenuated by dust, we wanted to be able to control the rate at which \lya~was attenuated with respect to the continuum, thus we applied a new free parameter, which we will call the clumpiness parameter (geometry parameter from F08a).  When we apply dust of a given optical depth to a continuum wavelength element, we multiply the flux by e$^{-\tau}$.  For the \lya~wavelength bin, we instead multiply by e$^{-q\tau}$, where q is the clumpiness parameter, which ranged from 0 -- 10.  When q is zero, \lya~photons are not attenuated by dust at all, and thus this represents the extreme clumpy ISM scenario.  When q is 10, resonance scattering works against the \lya~photons, and they suffer much greater extinction than continuum photons, which is the case in a homogeneous ISM.  Intermediate values of q represent intermediate geometries, but the important point is that for any value of q $<$ 1, the \lya~EW is being enhanced over that intrinsic to the stars.  We impress upon the reader that the dust is mixed in with the gas, thus it is the geometry of the gas which is dominating the \lya~radiative transfer, as \lya~photons are resonantly scattered.

We then took our model spectra, which were output by BC03 in units of L$_{\lambda}$, and converted them to f$_{\nu}$, redshifting them and attenuating their far-UV (FUV) with Madau (1995) IGM absorption in the process (see F08a for further details).  Although all of our \lya~flux is in one wavelength bin, it should not all be attenuated by the IGM, as in a real spectrum, about half of the flux would be on each side of the true wavelength of \lya.  Thus, when we applied the IGM, we only allowed it to attenuate half of the \lya~flux.  Lastly, we computed the bandpass averaged fluxes $\left<f_{\nu}\right>$ of each spectra through all of our observed filters (NB656, NB665, NB673, $B$, $V$, $i'$, $z'$, 3.6, 4.5, 5.8 and 8.0 $\mu$m), which gave us fluxes which we could directly compare to our candidate LAEs.

\section{Results}
In the time since F08a, we have updated our model grid and fitting program, thus we include the objects from F08a in our analysis, computing updated best-fit models for the four objects.  We have assigned them identifiers of CHa-1 through CHa-4, where ``C'' stands for CDF--S and ``Ha'' stands for the detection filter (NB656 or $H\alpha$ in this case).  CHa-1 and 2 are objects 1 and 2 from F08a, CHa-3 is object 4, and CHa-4 is object 6.  We also assign similar identifiers to our new objects, with CH8-1 and 2 representing the two candidate LAEs detected in NB665, and CS2-1 through CS2-8 representing the eight candidate LAEs detected in NB673.  We will use these identifies hereafter to represent our 14 candidate LAEs.  \tab{optmag}~\& \tab{irmag}~list each candidate by name, with their multi-wavelength magnitudes and their spectroscopic information when available.

\subsection{Model Fitting}

\subsubsection{Computing the Best-Fit Model}
We derived the best fit model to our observations via the method of $\chi^{2}$ fitting, where the model with the lowest reduced $\chi^{2}$ ($\chi^{2}_{r}$) was assigned to be the best fit for a particular galaxy.  $\chi^{2}_{r}$ is the $\chi^{2}$ divided by the number of degrees of freedom ($\nu$), which is defined as the number of constraints (data points) minus the number of free parameters.  While each object has numerous data points, we do not fit all of them to the models for various reasons.  First off, in order to separate model parameters from mass, we fit flux ratios (colors) rather than fluxes, computing the ratio of flux in each band to the $i'$-band.  We then derive the mass by scaling the best-fit model up to match the observations (using the detected $r'$, $i'$, $z'$, 3.6$\mu$m and 4.5$\mu$m bands).  Thus, while this means that we do not count mass as a free parameter, we likewise do not count the $i'$-band as a constraint.  Also, since our objects are narrow-band selected, we have a very good idea of their redshift, thus we do not include redshift as a free parameter.  For objects detected in NB656, NB665 and NB673, we assign the models a redshift of 4.399, 4.470 and 4.532 respectively (corresponding to the redshift of a \lya~emission line at the center of each filter).  As the $B$-band flux should remain undetected for our candidates if we have the redshift correct, the $B$-band data point does not add any physical information to our fits unless redshift is a free parameter, thus it is excluded.  The 5.8 and 8.0 $\mu$m fluxes of all of the candidates have a significantly less than 3 $\sigma$ significance, thus we do not include these bands in the fitting.

The FIREWORKS catalog also published the J, H and Ks band fluxes for objects from the VLT/ISAAC GOODS data.  As the FIREWORKS catalog was Ks-selected, it did not contain any of our objects.  However, given the importance of fitting the SED between observed  1 and 3.6 $\mu$m (rest 1800 - 6500 \AA), we examined the positions in the VLT/ISAAC J, H and Ks-band mosaics all of our 14 candidates.  In no cases were we able to definitively visually identify an object above the noise.  However, any information in this wavelength regime will help the model fitting.  The fluxes of our objects are high enough, and the images are deep enough, that including the upper limits should provide meaningful constraints (unlike IRAC 5.8 and 8.0 $\mu$m, where the upper limits are much higher than the expected object flux).  We thus include 3 $\sigma$ upper limits on the J, H and Ks bands during our model fitting.  We treat these upper limits (as well as the 3.6  and 4.5 $\mu$m points when they are undetected) in a manner similar to that of Nilsson et al. (2007).  Briefly, we set the flux of each object equal to its 3 $\sigma$ upper limit (the 3 $\sigma$ limiting magnitude of the images for the J, H and Ks bands, and the actual 3 $\sigma$ flux from the TFIT errors for the IRAC bands), with the 1 $\sigma$ flux error being equal to 33\% of its upper limit flux.  We then run our $\chi^{2}$ fitting code like normal, except that when an upper limit is compared to the model, a $\chi^{2}$ penalty is only assessed if the model flux is greater than the object (3 $\sigma$) flux.  In this manner models which exceed the upper limits are strongly disfavored.  

\subsubsection{Complicated Star Formation Histories}
While our range of star formation histories may accurately represent the SFHs of our objects, it may be more complicated.  Specifically, we would like to see if any of our objects are better fit by two bursts of star formation.  This could provide an alternate explanation to dust enhancement when an object has red colors, but a strong EW.  In this scenario, the strong \lya~EW could be from a relatively small fraction of young stars, with a large fraction of old stars causing red optical colors (while not contributing much to the rest-frame UV).  We follow the maximum age method of Papovich et al. (2001), where the young burst is allowed to vary in age, but the old burst is fixed to have occurred at z $\to \infty$, which at z $\sim$ 4.5 corresponds to t$_{pop}$ $\approx$ 1.4 Gyr.  We force both populations to have a burst SFR ($\tau_{SFH}$ = 10$^{5}$), and for simplicity we also assume that they have the same metallicity, dust extinction, and dust clumpiness.  Thus, the free parameters are:  t$_{pop}$ (young), Z, \dust, q and fraction of mass in old stars.  We chose a grid of allowable mass fractions, ranging from no mass in old stars, up to 99\%.

\subsubsection{Range of Allowable Parameters}
The process we have described above allows us to accurately compute the best-fit model to a given object SED.  While this may represent the true stellar population of the object, it is more critical to know the range of allowable parameter space, as in most cases there are many other models with $\chi^{2}_{r}$ near to the minimum value.  We chose to investigate these possible degeneracies via Monte Carlo simulations.  In each simulation, we vary the observed flux in each bandpass by a random amount proportional to the flux errors, recomputing the best fit.  We then run 7000 simulations, using different random numbers in each simulation, giving us 7000 best-fits per object.  These best-fits show us how the uncertainties in our observations can lead to changes in the best-fit models.  In cases where the best-fits do not change much, we can assume that the original best-fit model well constrains the stellar population in the given object.  However, in cases where the Monte Carlo results vary widely, our models are not able to constrain the objects very well.

\subsection{NB656 Detected Objects}
\begin{figure}
\epsscale{1.75}
\begin{center}\plottwo{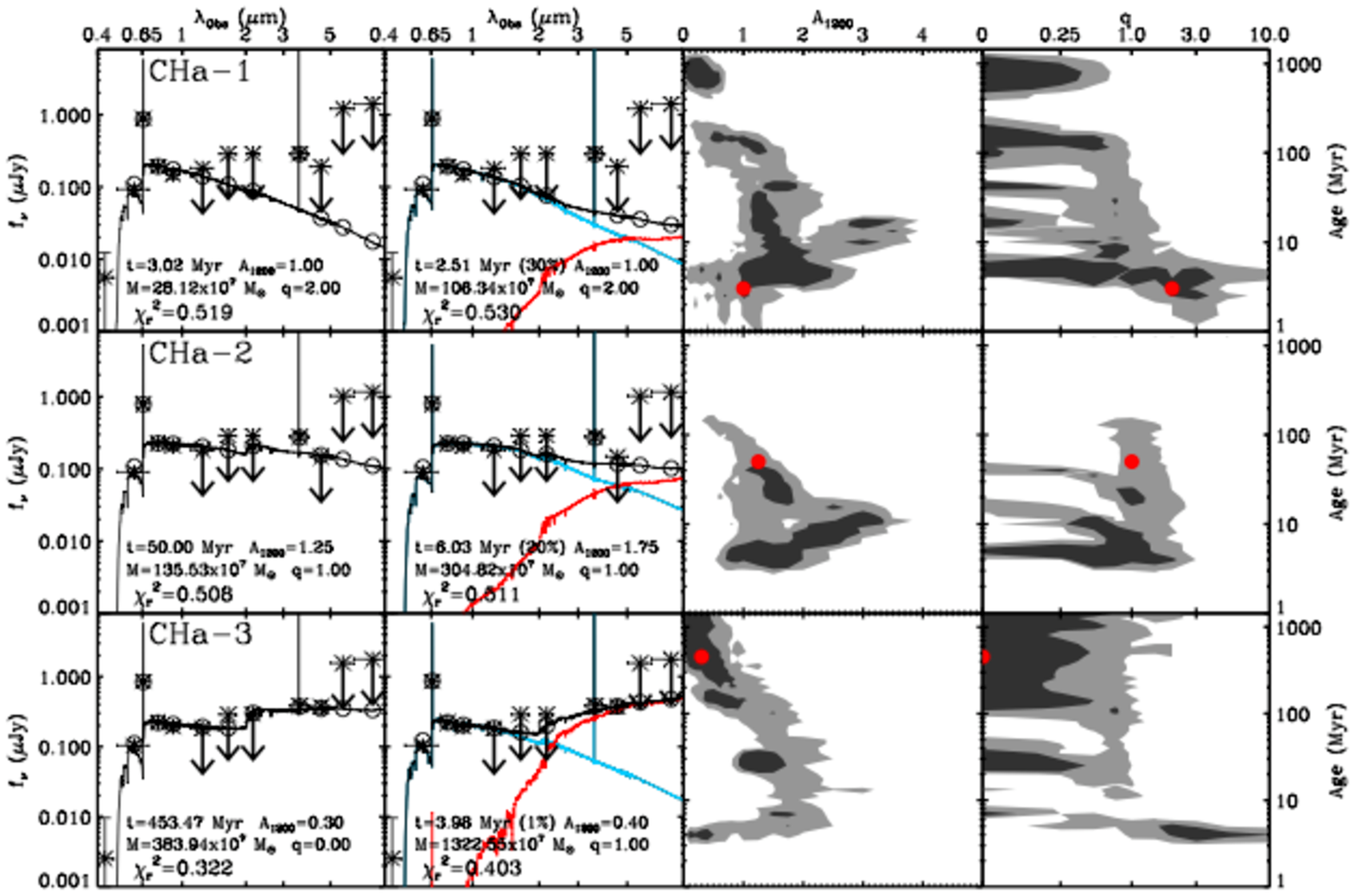}{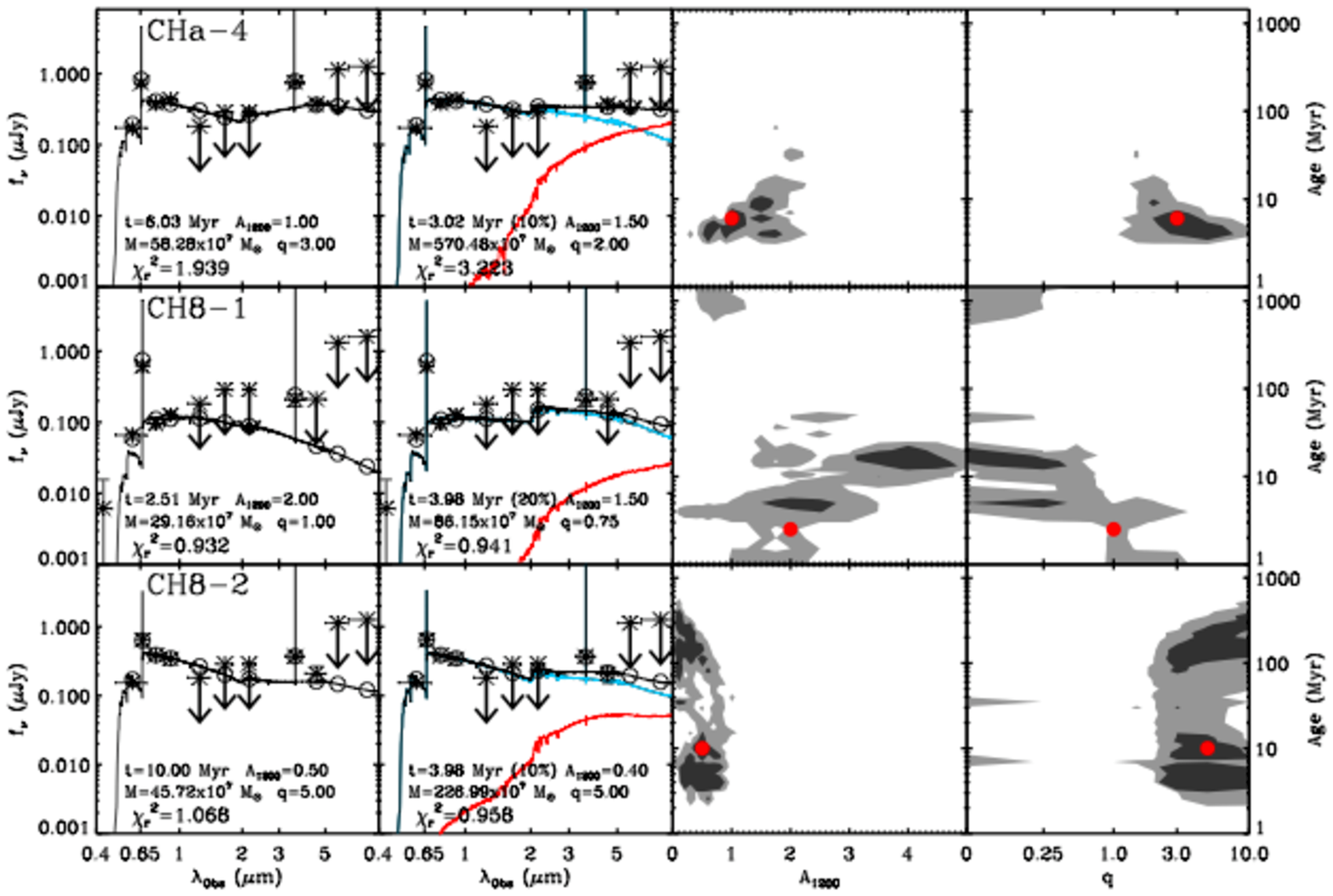}\end{center}
\caption{ The first column contains the best-fit single population models to the first six of our candidate LAEs, with their observed SEDs over-plotted.  The open circles represent the bandpass-averaged fluxes of the models, highlighting the $H\alpha$ dominated 3.6 $\mu$m flux of the younger objects.  The second column has the best-fit two-burst models to our objects.  Objects with no mass in old stars can still have a worse fit than those in \fig{sed1}, as they are forced to have a burst SFH.  The third and fourth columns contain the Monte Carlo simulation results (for single-population models), with the best-fit point in red.  The dark and grey shaded areas represent the 68\% and 95\% confidence intervals respectively.}\label{sed1}
\end{figure}

\begin{figure}
\epsscale{2.0}
\begin{center}\plottwo{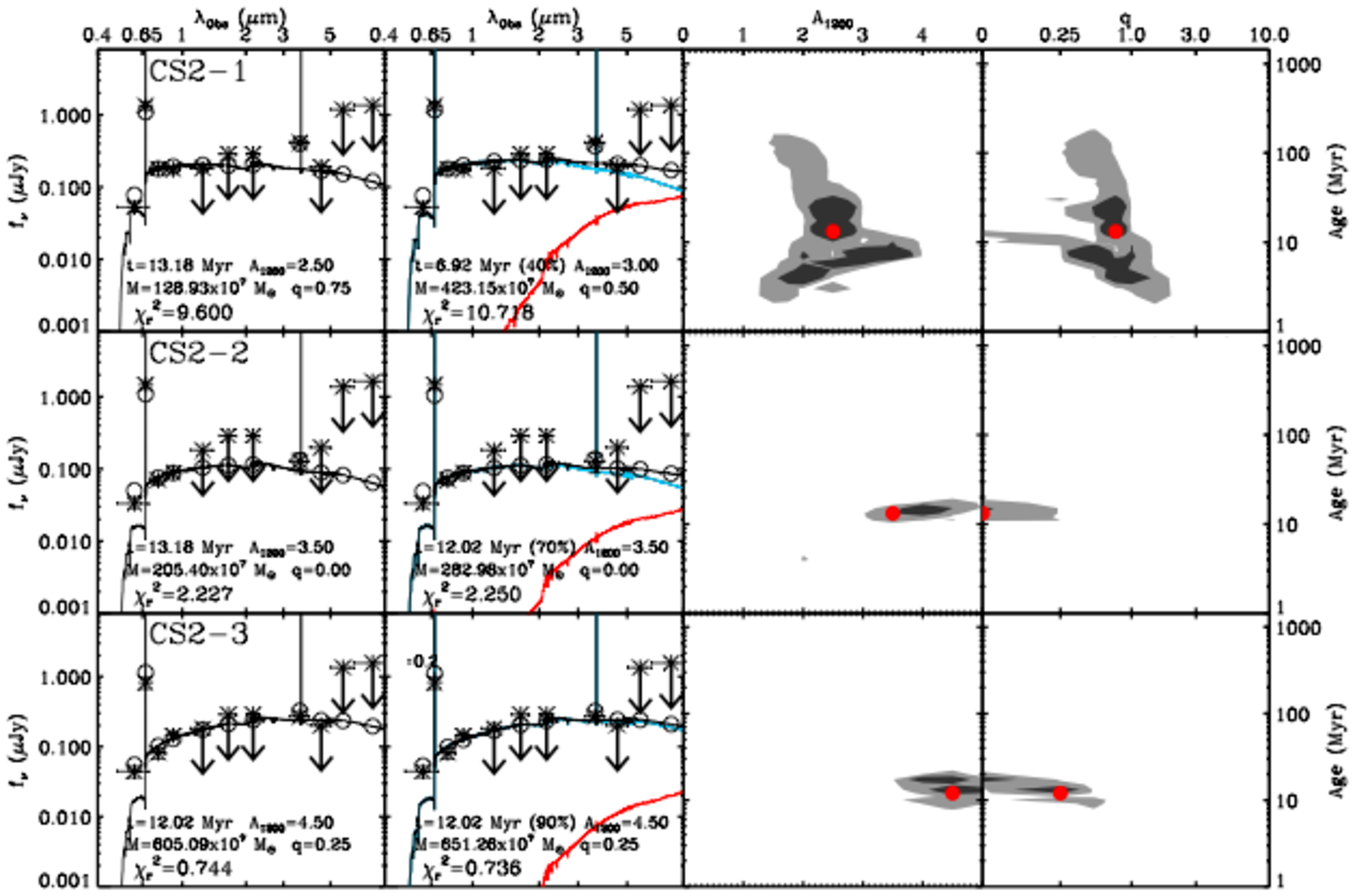}{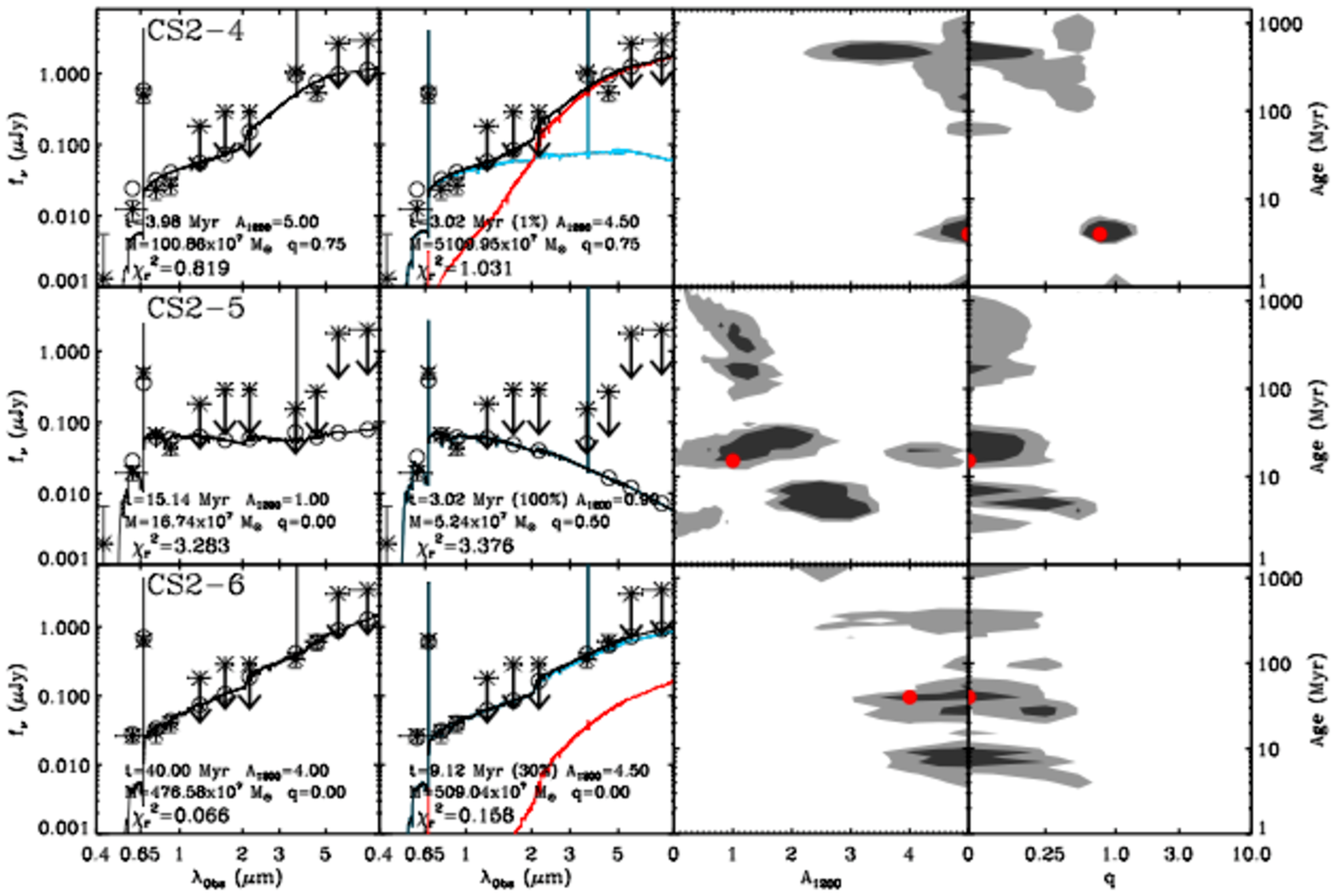}\end{center}
\caption{Same as \fig{sed1}, for the next six objects.}\label{sed2}
\end{figure}

\begin{figure}
\epsscale{0.90}
\begin{center}\plotone{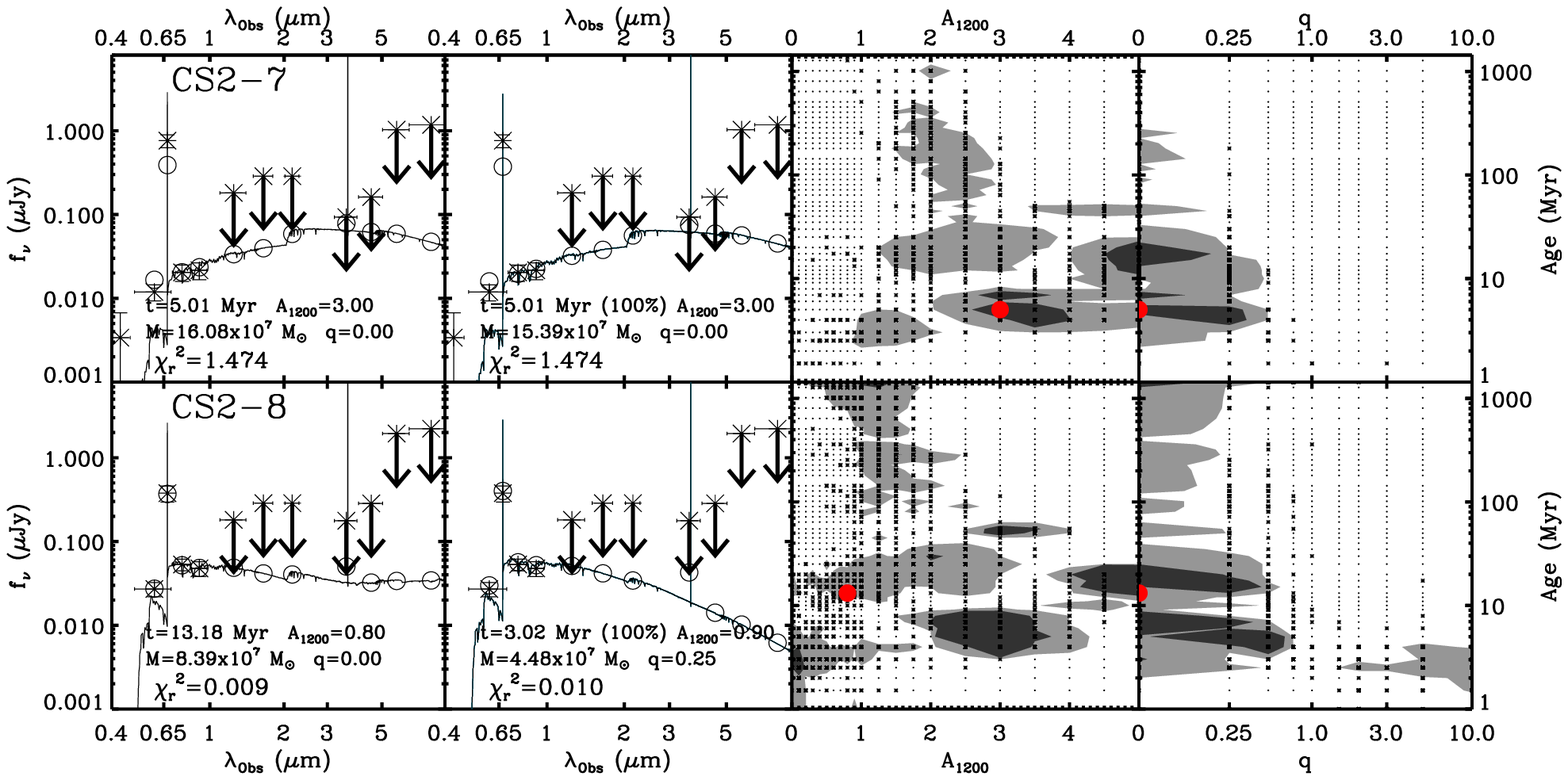}\end{center}
\caption{ The same as \fig{sed1} \& \fig{sed2}, for the last two objects.}\label{sed3}
\end{figure}

We will now go through a detailed discussion of the results for each object.  These results are shown in \fig{sed1}, \fig{sed2}~and \fig{sed3}.  These figures are arranged as follows.  Each row corresponds to one object.  The first column displays the best-fit single population model to the data, listing out some of the best-fit results.  The second column is similar, except that it displays the best-fit two-burst model.  The third and fourth columns display the results of the Monte Carlo analysis for the single-population models, with the third column displaying t$_{pop}$ vs. \dust, and the fourth column displaying t$_{pop}$ vs. q (clumpiness).  We tabulate the best-fit single-population and two-burst models in \tab{singlefit}~and \tab{tbfit}~respectively.

\subsubsection{CHa-1}
The best-fit model for object CHa-1 (object 1 from F08a) is very young and relatively low-mass, at 3 Myr and 2.8 $\times$ 10$^{8}$ $M$\sol.  This model has 1 magnitude of dust extinguishing \lya~somewhat more than the continuum (q = 2.0).  The metallicity is best-fit by 20\% of Solar, higher than the fixed metallicity value from F08a, and the SFH is best-fit by continuous star formation.  These parameters are very similar to those from F08a.  We attribute the small differences in age, dust and q to the different treatment of the IRAC upper limits in the current work, as well as they ability of metallicity to vary.  The best-fit model has a rest-frame EW of 81 \AA, which is low for 3 Myr, but as q = 2, the \lya~flux is being attenuated by some of the dust.  This is significantly less than the computed EW from the object's narrowband excess of 190 $\pm$ 47 \AA.  We attribute this difference to being due to the uncertainty in where the \lya~line falls in both the narrowband and the broadband filters, thus the F606W flux may not be indicative of the true continuum flux due to intervening IGM absorption.  See \S 5.1 for further discussion on this matter.  The best-fit two-burst model is also a good fit, containing 70\% of its mass in old stars, for a mass-weighted age of 1005 Myr.  This two-burst model is an example of an object which exhibits strong \lya~emission without the need for an extremely clumpy ISM, even though its mass is dominated by older stars.  The results of the Monte Carlo simulations show that while the best-fit age is young, there are islands of allowable parameter space as old as several hundred Myr, implying that the best-fit model is not unique in explaining the observed SED of this object (also explaining why the two-burst model is a good fit).  To fit this object, the older models have correspondingly less dust (as the older age will redden the colors, so dust is less necessary), and a smaller q (as the age goes up, q will have to go down to keep the \lya~flux constant).  The 68\% confidence intervals span a somewhat larger region of allowed parameter space than those for this object in F08a, which we attribute again to the different treatment of the IRAC upper limits, and also the addition of the NIR upper limits.
\subsubsection{CHa-2}
CHa-2 (object 2 from F08a) is older, at 50 Myr, and more massive, at 1.4 $\times$ 10$^{9}$ $M$\sol.  It is best-fit by \dust~= 1.25, although the clumpiness (q) value of 1 implies that the dust does not affect the value of the EW.  The metallicity is 2\% of Solar, consistent with the fixed value in F08a, and this object has an exponentially decaying SFH with $\tau_{SFH}$ = 10$^{8}$ yr.  The model EW of 76 \AA~is near what we would expect for 50 Myr, thus dust enhancement or extinction of the line is not necessary.  This model is both older and higher mass than in F08a, and in this case the new treatment has resulted in a much better fit, with a higher \dust~and a lower q in the current work.  The best-fit two-burst model is a slightly worse fit, but it is significantly different as it has 80\% of its mass in old stars, and its total mass is a factor of 2.3X more.  It still has q = 1, thus the age of its young component is very young in order to keep the \lya~flux up.  The age of this object is better constrained than CHa-1, as nearly all of the Monte Carlo simulations result in t$_{pop}$ $\lesssim$ 100 Myr.  However, large ranges of both dust and q are allowed, although some amount of dust and a value of q $\leq$ 3 do seem to be strongly favored.
\subsubsection{CHa-3}
We were very interested in the updated results for CHa-3 (object 4 in F08a), as it was by far the most interesting object in F08a, showing definitive signs of dust enhancement of the \lya~EW, with a best-fit of 800 Myr, 6.5 $\times$ 10$^{9}$ $M$\sol, \dust = 0.4 and q = 0.  The updated results for this object confirm those from F08a, as a large amount of dust enhancement appears to be occurring, allowing this object to exhibit a decently strong \lya~EW ($\sim$ 100 \AA) at an old age.  The updated best-fit age of 454 Myr is less than that from F08a (and it is less massive at 3.8 $\times$ 10$^{9}$ $M$\sol), yet it is still best-fit by q = 0 with \dust = 0.3 mag.  It is still fit by continuous star formation, but now that metallicity is allowed to vary, the new, higher best-fit value of Z = 0.2$Z$\sol~also explains the younger age (the higher metallicity will result in redder colors, thus the best-fit age is lowered from 800 Myr to 454 Myr to cancel out this effect).  The two-burst best-fit is a slightly worse fit, but it can also explain this object's SED, with 99\% of the mass in old stars (with the other 1\% of the mass from a 4 Myr old burst giving a mass-weighted age of $\sim$ 1420 Myr), and the q of 1 (and \dust = 0.4) shows that while the line is not actually being enhanced, some amount of clumpiness is still required in order to keep the \lya~flux from being completely extinguished.  The Monte Carlo results show that the age of this object is strongly favored to be $>$ 100 Myr, although there are a few small 68\% confidence islands at younger ages.  This object is well constrained to contain a very clumpy ISM, as nearly all of the simulations result in q $\leq$ 1 (with a small island at q $\sim$ 3).  These results further confirm our previous interpretation that this object represents a new class of LAE: one that is old yet still exhibits strong \lya~emission due to a clumpy ISM enhancing the escape fraction of \lya~photons over continuum photons (although the Monte Carlo results show that there is a slight chance that this object is younger, with dust heavily extinguishing the \lya~flux).
\subsubsection{CHa-4}
The last of the objects from F08a, CHa-4 (object 6), again shows similar results to F08a, with a young age (6 Myr) and low mass (5.8 $\times$ 10$^{8}$ $M$\sol).  The dust extinction is less, at \dust = 1.0 (vs. 1.8 from F08a), yet the q value has increased to keep a similar extinction of the \lya~flux (this model has a low EW of only 27 \AA, consistent with our observed value of 56 \AA).  However, this model is not well constrained, as evidenced by its relatively high $\chi^{2}_{r}$ value, and the violation of the J-band upper limit.  The shape of the observed SED is interesting, as both the z' and 4.5$\mu$m flux are higher than the NIR upper limits.  This could mean that two-burst fit could produce a more consistent model, as the optical/NIR fluxes could be due to a young component, with the IRAC fluxes due to a significant old component.  This is in fact what the best-fit two-burst model shows, with only 10\% of the stellar mass in young (3 Myr) stars.  However, the $\chi^{2}_{r}$ of this model is even higher, thus this object awaits deeper NIR data to fully analyze its SED.

\subsection{NB665 Detected Objects}
\subsubsection{CH8-1}
The first of our new sample, CH8-1, has a best-fit stellar population that is young and dusty, with its best-fit model of t$_{pop}$ = 2.5 Myr, 2.9 $\times$ 10$^{8}$ $M$\sol, \dust~= 2.0 and q = 1.  The best-fit \lya~EW is somewhat high, at $\sim$ 130 \AA, and appears to be unaffected by dust.  The two-burst model also shows a good fit, with only 20\% of its mass in young stars, and a small amount of dust enhancement helping to keep the line strong (q = 0.75).  This two-burst model has a mass three times greater than the single population model, but it's metallicity is reduced by a factor of five.  Nearly all of the simulations result in best-fit models with young ages, with q $<$ 1 and a good amount of dust, although there is a small island at very old age, with some dust and q = 0.  This uncertainty likely derives from the undetected 4.5 $\mu$m flux, as the 3.6 - 4.5 $\mu$m color constraints the amount of $H\alpha$ emission, which should only be strong in a young, star forming galaxy.  We will return to this issue in \S 5.3.1.
\subsubsection{CH8-2}
This object is best-fit by a young (10 Myr), medium (for LAEs) mass (4.6 $\times$ 10$^{8}$ $M$\sol) stellar population.  Normally a galaxy this young would show very strong \lya~emission, however this object is best-fit by \dust~= 0.5 mag with the highest value of the clumpiness parameter (meaning it has the most homogeneous ISM) in our sample of q = 5.  The best-fit two burst model is actually a slightly better fit, with only 10\% of its mass in a young (4 Myr) population, for a mass-weighted age of 1291 Myr, and six times more massive than the single population model.  This model also shows significant attenuation of the \lya~flux, with q = 5.  Both \dust~and q appear to be very well constrained to be low and high respectively, however the age does have a few allowed regions, one around the best-fit at 10 Myr, and a few more at higher age.

\subsection{NB643 Detected Objects}
\subsubsection{CS2-1}
CS2-1 has the lowest quality fit of our sample, which appears to mainly come from the inability of the model to account for both the V-band and narrowband flux.  The best-fit model is somewhat young, although higher mass, at 13.2 Myr and 1.3 $\times$ 10$^{9}$ $M$\sol, with a lot of dust (2.5 mag) extinguishing the continuum slightly more than \lya~(q = 0.75), thus the \lya~EW is being slightly enhanced.  The best-fit two-burst model is an even worse fit.  The allowed parameter space is of moderate size, with the age constrained to be $<$ 100 Myr, some amount of dust, and q likely $\leq$ 2.  However, given the poor quality of the best-fit, we do not lend much weight to these results.
\subsubsection{CS2-2}
This object has a similar age as CS2-1, with a higher mass of 2.0 $\times$ 10$^{9}$ $M$\sol.  A stellar population of this age with this object's best-fit SFH ($\tau_{SFH}$ = 10$^{6}$) will have a relatively low \lya~EW.  However, the EW of this best-fit model is 400 \AA, implying significant enhancement of the EW, coming from the 3.5 mag of dust extinction from a very clumpy (q = 0) ISM.  This object has a very large narrowband excess, due mostly to the faintness of its V-band flux, which is why the best-fit model needs a high EW.  However, the i' - z' color is very red, thus the object is better fit by a 13 Myr population with clumpy dust responsible for the red colors and high EW, rather than a very young population.  The two-burst fit is also a good fit, with 30\% of its mass in old stars.  The two-burst model also shows dust enhancement, with the same \dust~and q as the single population model.  The best-fit parameters are very well constrained to have an age $\sim$ 10 Myr, and a lot of clumpy dust.  This is another object which we can add to our growing list of those that require dust enhancement of the \lya~EW to fully fit their observed SED.
\subsubsection{CS2-3}
CS2-3 has a very similar population to that of CS2-2, although this object has a substantially higher quality fit.  Its best fit parameters of 12 Myr, 6 $\times$ 10$^{9}$ $M$\sol, \dust~= 4.5 mag and q = 0.25 imply that a significant amount of dust enhancement is responsible for its \lya~EW of 341 \AA.  The best-fit two-burst model shows a nearly identical population, with 10\% of its slightly greater mass in old stars.  The physical parameters of this object also appear to be very well constrained, as shown by the Monte Carlo simulation results.
\subsubsection{CS2-4}
The observed SED of CS2-4 is very red, yet the 3.6 - 4.5 $\mu$m color favors a young population with red colors due to dust extinction, thus this object is best-fit by a 4 Myr old population, with a mass of 10$^{9}$ $M$\sol, and 5 mag of dust extinction, with some amount of dust enhancement occurring (q = 0.75).  The two-burst model has a slightly poorer fit, but it is a very different population with a mass-weighted age of 1420 Myr and a mass of $\sim$ 5 $\times$ 10$^{10}$ $M$\sol, as well as significant dust enhancement.  While both q and \dust~are relatively well constrained, the simulations fall into two different age ranges; one at the best-fit value of $\sim$ 5 Myr, and another, larger region at $\sim$ 500 Myr.  This island of older best-fit models requires q = 0.  Thus, we conclude that this object could be a young population with a small amount of dust enhancement, or a very old population with a large amount of dust enhancement.
\subsubsection{CS2-5}
This object has a best-fit age of 15 Myr, with a mass of 1.7 $\times$ 10$^{8}$ $M$\sol.  This object also requires dust enhancement in its best-fit model, with \dust~= 1.0 mag and q = 0. Both the best-fit single population and two-burst models have similar fit qualities, although both are relatively poor fits.  While q is relatively well constrained to be small, both \dust~and t$_{pop}$ span a wide region of allowed parameter space.  This object is undetected in all {\it Spitzer} bands, thus these results may not be constraining.
\subsubsection{CS2-6}
CS2-6 has one of the highest quality fits in our sample, and it too exhibits evidence for dust enhancement of the \lya~EW.  The very red SED is best-fit by a 40 Myr, 4.8 $\times$ 10$^{9}$ $M$\sol~population, with all 4.0 mag of dust extinction affecting the continuum only, significantly enhancing the \lya~EW (1000 \AA).  The best-fit two-burst model has a worse fit quality, with a similar combined population.  This is allowed as the undetected NIR fluxes do not constrain the amplitude of the 4000 \AA~continuum break (see \S 5.3.1).  Both dust and q are relatively well constrained.  The age is somewhat constrained to be $\leq$ 40 Myr, but there are a few small allowed regions at higher age.
\subsubsection{CS2-7}
CS2-8 has a best-fit model of 5 Myr and 1.6 $\times$ 10$^{8}$ $M$\sol, with dust enhancing its already high EW (\dust~= 3.0; q = 0).  Many values of dust appear to be allowed, with \dust~= 3 to 4 mag favored.  Dust enhancement looks to be very favored in this object, as nearly all simulations fall at q $\leq$ 0.5.  The best-fit two-burst model is identical to the single population model.  However, this object is undetected in all IRAC bands, thus the best-fit models are not very well constrained.
\subsubsection{CS2-8}
Lastly, while CS2-8 has the lowest $\chi^{2}_{r}$ of all of our objects, this is partly due to the fact that we again have no detections redward of the z'-band, thus while we report our results for this object, they only represent a possible population for this object; deeper NIR and IR data are needed to fully constrain this object.  Nonetheless, this object also shows dust enhancement of the \lya~EW, with \dust~= 0.8 mag and q = 0 in its best-fit 13.2 Myr old, 8.4 $\times$ 10$^{7}$ $M$\sol~population.  However, with the exception of q, the best-fit parameters are very poorly constrained, as evidenced by the simulation results.

\section{Discussion}

\subsection{Equivalent Width Distribution}
\begin{figure}
\epsscale{1.0}
\begin{center}\plotone{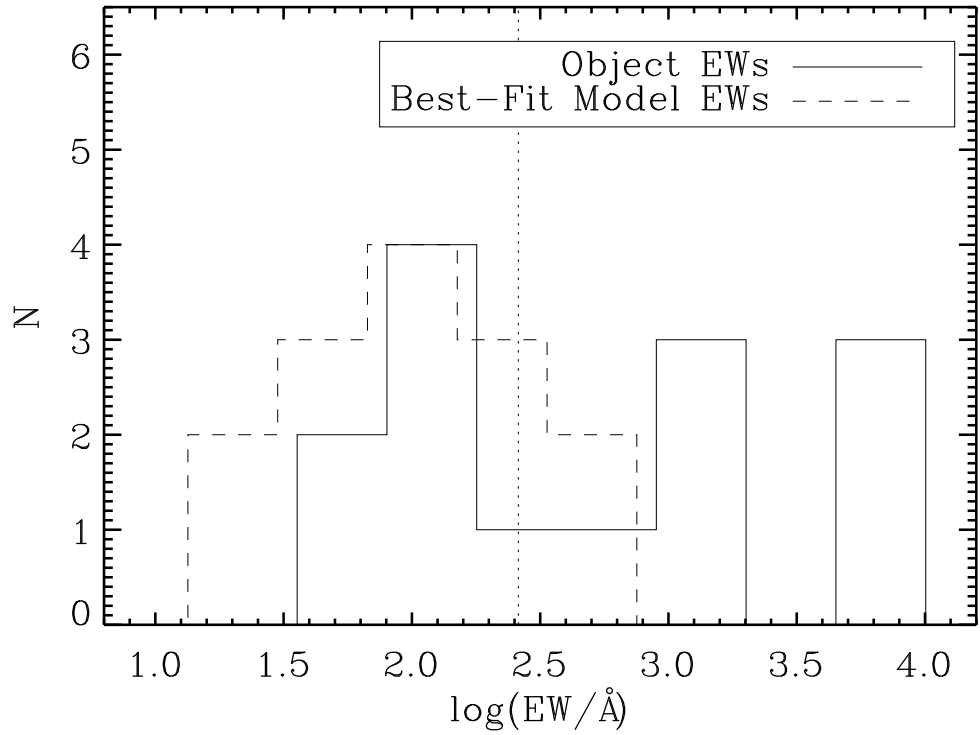}\end{center}
\caption{The rest-frame EWs as measured from our narrowband and continuum (F606W) observations are plotted in the solid histogram, with the rest-frame EWs from the best-fit models plotted in the dashed histogram.  The dotted line denotes an EW of 255 \AA, which is the maximum EW allowed by a Z = 0.02~$Z$\sol~stellar population (at 1 Myr).  Eight (five) objects (best-fit models) have computed EWs $>$ 255 \AA, implying that other effects (i.e. dust enhancement etc.) are in play.  We are inclined to trust the model EWs over the computed EWs of the objects, as the object EWs suffer from the uncertainty of where the line falls in the narrowband and F606W filters.}\label{ews}
\end{figure}

In \fig{ews}~we plot both the distribution of EWs derived from the computed narrow-band excess of our objects (see Finkelstein et al. 2007 for this calculation), and the EWs from the best-fit models.  While both distributions peak at a rest-frame EW of $\sim$ 100 \AA, the object EWs show a larger number of high EWs, and vice versa for the model EWs.  As we alluded to in \S 4, we believe that this is due to the inherent uncertainty in calculating EWs from the observed fluxes.  This calculation relies on a measurement of the line-flux to continuum-flux ratio, and due to intervening IGM material we lack a true measurement of the continuum flux in the regime surrounding the line.  Examining the V-band data points in \fig{sed1}, \fig{sed2}~and \fig{sed3}, one can see that the bandpass is nearly all blueward of the \lya~line, thus it does not contain a true measurement of the intrinsic continuum flux.

However, the EWs of our models are computed in a more accurate fashion, as we have the entire model spectrum at our disposal.  Our model EWs are calculated using:
\begin{equation}
EW_{rest,model} = \frac{f (\lambda_{Ly\alpha})}{f (\lambda_{Ly\alpha+1})}*(\lambda_{Ly\alpha+1} - \lambda_{Ly\alpha})*(1 + z)^{-1}
\end{equation}
where $\lambda_{Ly\alpha}$ is the wavelength position of the \lya~line (1215 \AA~in the model; the entire line is concentrated in this bin), and $\lambda_{Ly\alpha+1}$ is the wavelength element immediately to the right of \lya~(1217 \AA).  In this fashion we obtain a true measurement of the stellar continuum near the line, and thus, at least when the quality of the fit is acceptable, we consider the EWs from the best-fit models as being more indicative of the true value.

For our purposes, we define a ``normal'' stellar population as one with a continuous SFH, Salpeter IMF, and Z = .02$Z$\sol~(i.e., no dust, top-heavy IMF etc. causing high EWs).  This population has a maximum \lya~EW of 255 \AA~at 1 Myr, asymptoting to a constant value of 76 \AA~by 100 Myr.  We can examine the importance of q in our sample by comparing our best-fit model EWs to this normal stellar population.  We find that 11 of our best-fit model EWs are higher than 76 \AA, meaning that they have to be younger than 100 Myr, which is consistent with our results.  However, we also find that five of these 11 have best-fit model EWs $>$ 255 \AA, which means that dust enhancement is likely to be occurring to explain their EWs.  This is confirmed by our results, as four of these five have q = 0, with the other having q = 0.25, thus showing that a clumpy ISM is significantly enhancing their \lya~EW.

\subsection{Model Parameter Distribution}
\subsubsection{Best-Fit Models}
\begin{figure}
\epsscale{1.2}
\plottwo{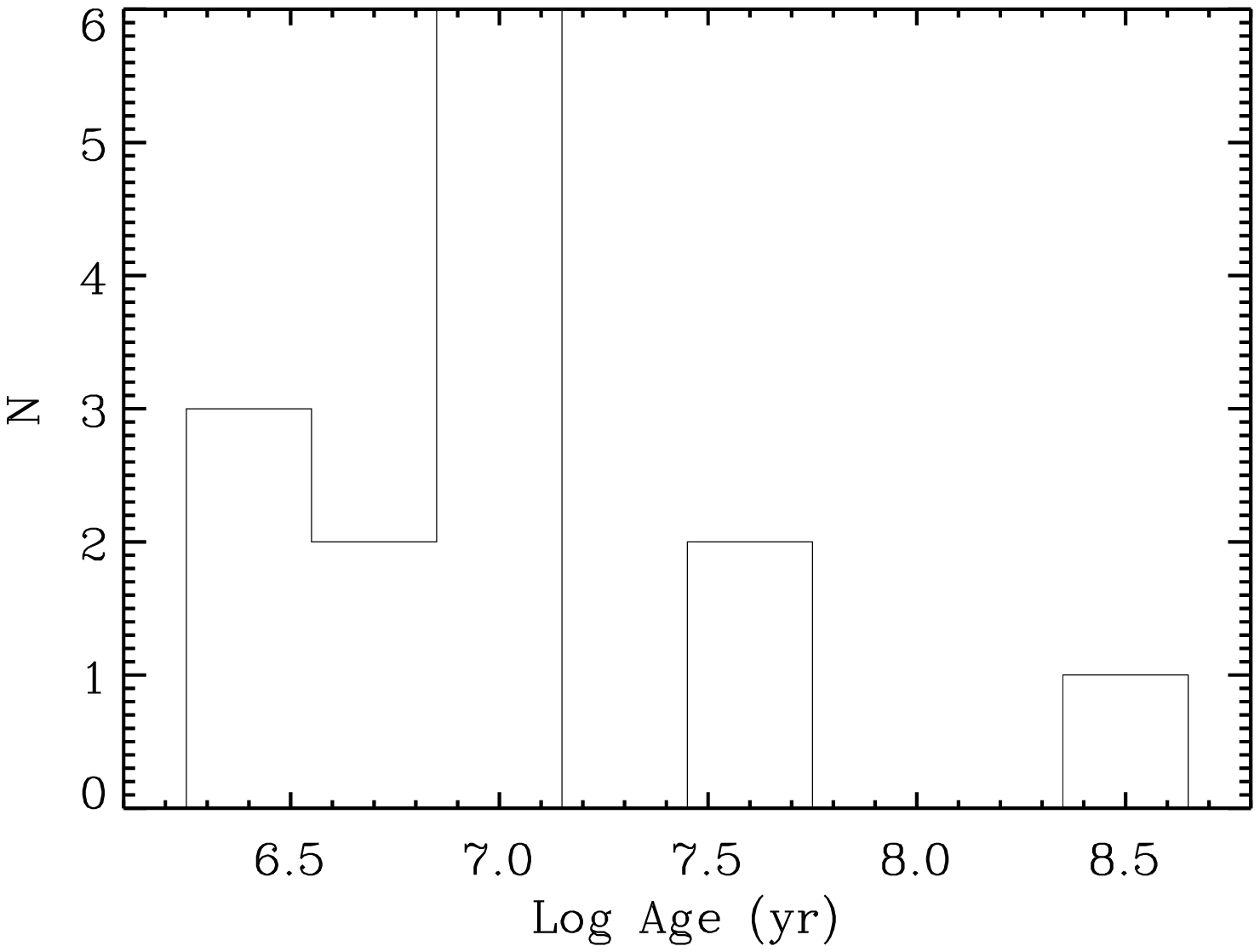}{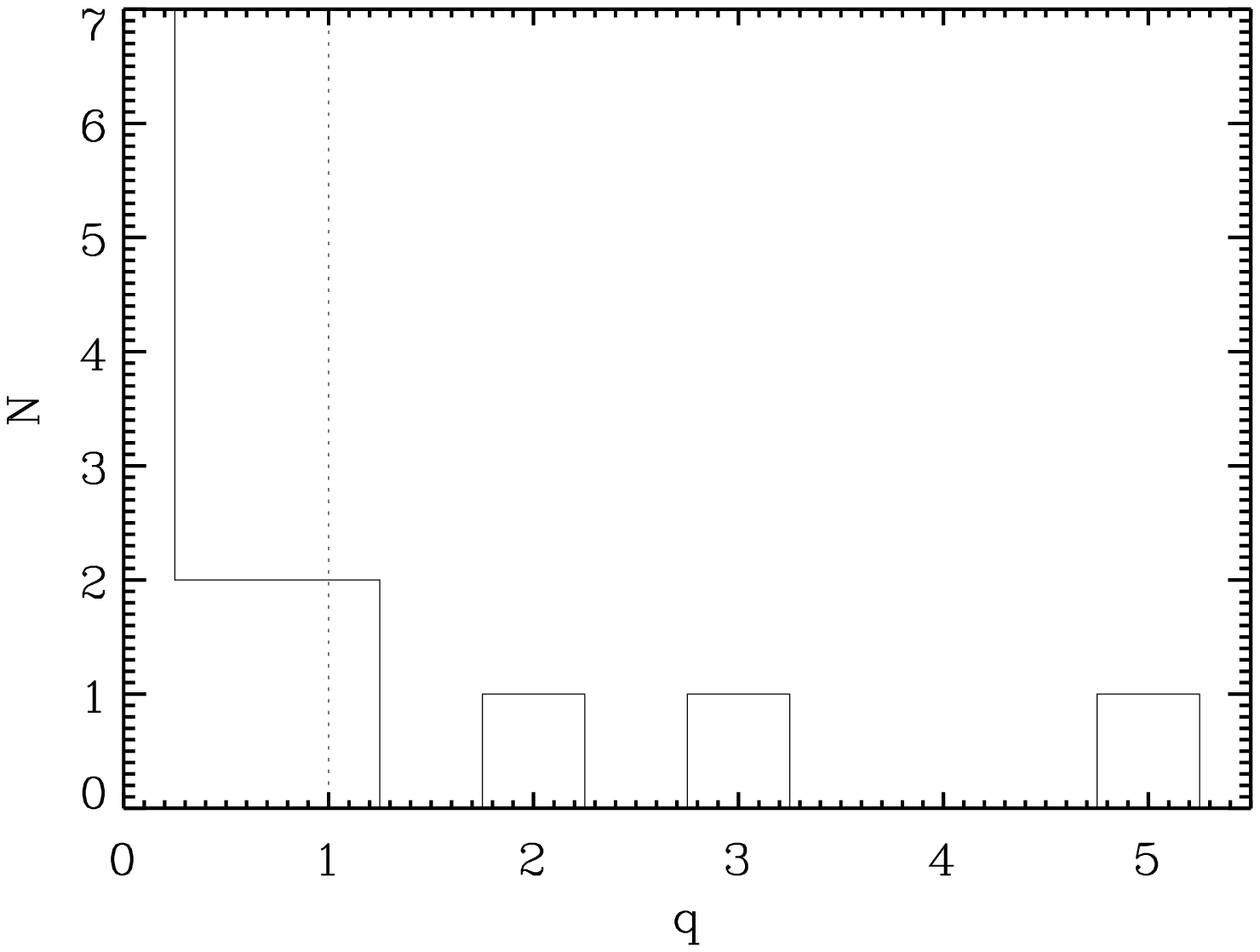}
\caption{The distribution of best-fit stellar population ages and clumpiness parameters from our single component models.  The majority of our objects have young ages, and a majority of our sample also requires clumpy dust to enhance its \lya~EW.}\label{fithist}
\end{figure}

In \fig{fithist}a we show the distribution of best-fit ages from our models.  In F08a, out of four galaxies, we found that three had ages of near to 10$^{6}$ yr, with one near 8 $\times$ 10$^{8}$ yr.  In our current study, we see that most of our objects have ages $\lesssim$ 10 Myr, with three out of the 14 objects showing ages $\geq$ 40 Myr.  However, we still have a large age range where we find no objects, from 50 - 450 Myr.  There are many scenarios one could think of to fit this distribution.  Perhaps the distribution of ages is continuous, and we just need a larger sample to fill in the gaps.  One intriguing possibility is that this distribution is true, i.e. we see \lya~galaxies when they are very young (and still see a few out to a few 10's of Myr), but then some physical mechanism is blocking the \lya~emission until much older age.  Before the first generation of stars dies, \lya~photons may find it easier to escape (especially if the initial dust is clumpy).  After a few 10's of Myr, stars will die and these young galaxies will be actively forming dust.  As the dust gets thicker, the amount of \lya~escaping will be reduced.  However, after some period of time, the stars may begin to ``punch holes'' through the dust, resulting in a clumpy (or hole-y) ISM that can enhance the observed EW, creating an object like CHa-3.  

In \fig{fithist}b we show the distribution of the best-fit clumpiness parameter.  Stunningly, we find that 64\% (9/14) of our objects appear to require an ISM dominated by a clumpy geometry (q $<$ 1) in order to explain their observed SEDs.  A few objects have q $\geq$ 3, indicating a more homogeneous ISM, although both of these objects have interesting fits, mostly due to their strange ``V''-shaped SEDs.

From \tab{singlefit}~we see that our derived masses span a large range, from 8.4 $\times$ 10$^{7}$ to 6 $\times$ 10$^{9}$ $M$\sol, with half of our sample at or over 10$^{9}$ $M$\sol.  These results are similar to those seen in past studies.  We (Finkelstein et al. 2007) have previously found a mass range of 10$^{8}$ to 2 $\times$ 10$^{9}$ $M$\sol~in a sample of 21 LAEs from ground-based photometry at z $\sim$ 4.5, with an age range of 4 -- 200 Myr.  Pirzkal et al. (2007) found a mass range of 7 $\times$ 10$^{6}$ to 1.4 $\times$ 10$^{9}$ $M$\sol, and an age range of 1 -- 20 Myr at z $\sim$ 5.  These objects likely show a smaller maximum age and mass as they were detected via grism spectroscopy in the {\it Hubble} Ultra Deep Field, thus these authors probed deeper into the luminosity function.  Gawiser et al. (2006) stacked a large number of LAEs at z $\sim$ 3.1, and found an average age of 90 Myr, with a mass of 5 $\times$ 10$^{8}$ $M$\sol.  Lai et al (2007; 2008) performed two studies, one at z $\sim$ 5.7, and one at 3.1.  At z $\sim$ 5.7, they found a mass range of 1.4 $\times$ 10$^{9}$ to 1.4 $\times$ 10$^{10}$, although this high upper mass is likely due to these authors sampling the high end of the mass function, as they reported results from IRAC detected objects.  These objects had an age range of 5 - 700 Myr.  At z $\sim$ 3.1, they separated their samples into IRAC undetected and detected stacks, and found 200 Myr and 3 $\times$ 10$^{8}$ $M$\sol~for the undetected stack, and 1 Gyr and 10$^{10}$ $M$\sol~for the IRAC detected stack.  Chary et al. (2005) found one LAE at z = 6.56, with a best-fit stellar population of 5 Myr and 8.4 $\times$ 10$^{8}$ $M$\sol.  Overall, we have found a broad range of ages and masses, but they appear consistent with these previous results.

While our detection of dust is very interesting, many of these other studies have found evidence for dust extinction as well.  While Lai et al. (2008) and Gawiser et al. (2006) do not find dust in their stacking analyses, Chary et al. (2005), Pirzkal et al. (2007) and Lai et al. (2007) do find dust in their analyses of individual objects, with A$_{V}$ as high as 1.3 (A$_{1200}$ is roughly a factor of 4 greater than A$_{V}$).  This shows an interesting trend, in that studies which analyze objects separately seem to detect dust extinction, while those that stack fluxes do not.  To see if our objects followed this trend, we did a quick test by averaging the fluxes in each band from our best-fit models, and then re-fitting this ``stack'' of our objects.  We found that our sample has an average stellar population with t$_{pop}$ = 6 Myr, 4.7 $\times$ 10$^{8}$ $M$\sol, \dust~= 2.0, q = 1 and EW $\sim$ 90 \AA, with a 99\% confidence level of containing some measure of dust.  This average population appears to be indicative of the average properties of our sample, including the detection of dust. 

\subsubsection{Most-Likely Models}
Although we find many objects with q $<$ 1, which allows \lya~emission at older age, we find that still only one object is best-fit by an old ($>$ 100 Myr) stellar population.  However, taking the Monte Carlo results into account, we can assign some objects more likely ages and q's, as the Monte Carlo simulations illuminate the likelihood range of these parameters.  For those objects where their best-fit lies in the largest 68\% confidence region, we leave their age and q fixed.  However, if this is not true, we assign them the age and q from the center of the largest 68\% confidence region from the Monte Carlo age vs. q contour plots.  We do this for: CHa-2, CH8-1, CS2-4, CS2-6 and CS2-8.  \fig{fithist2}a and \fig{fithist2}b show updated histograms using the new ages and q's for these five objects.

While the q distribution is mostly unchanged (although now 71\% of objects have q $<$ 1), the slight change in the distribution of ages has important consequences.  First, both CHa-2 and CS2-6, which had ages of 40 and 50 Myr respectively, are much more likely to have ages of $<$ 10 Myr.  Secondly, CS2-4, which had a best-fit age of 6 Myr with only a little dust enhancement, has a most likely age of 500 Myr with lots of dust enhancement.  Thus, our distribution has significantly changed, with 12 of our objects at $\leq$ 15 Myr, and 2 objects at $\geq$ 450 Myr, both with significant amounts of dust enhancement.  As the Monte Carlo results highlight how the observational errors can lead to uncertainties in the model fitting, we regard these most-likely models as the true results of our work.  These new results show an even greater bimodality than the best-fit results, suggesting a possible recurrence of the \lya~emitting phase in these galaxies (e.g., Shapley et al. 2001; Malhotra \& Rhoads 2002; Lai et al. 2008).

As another test, we took the Monte Carlo results and computed the probability of each object being fit by each age in our age grid, normalized to the number of simulations (i.e. so that the total probability is normalized to 1).  We could then look at probability curves for each object.  In order to see the probability distribution of age across our whole sample, we averaged all of these curves for each object, creating \fig{prob}.  This figure verifies the by-eye results of \fig{fithist2}a, in that we see two distinct peaks in age for our LAEs, at 4 Myr and 400 Myr.

We also examined these most-likely models in order to see how much the mass changed for these objects vs. their best-fit results.  As one would expect, the difference was large for CS2-4, which has a most-likely mass of 5.0 $\times$ 10$^{10}$ $M$\sol, which is 50X larger than the mass derived for this object from the best-fit young stellar population.  The differences in mass for CHa-2, CH8-1, CS2-6 and CS2-8 were smaller, with their most-likely masses being 5.3 $\times$ 10$^{8}$, 1.6 $\times$ 10$^{9}$, 9.5 $\times$ 10$^{8}$ and 3.3 $\times$ 10$^{8}$ $M$\sol~respectively.  This brings our upper end up to 5.0 $\times$ 10$^{10}$ $M$\sol, consistent with the upper mass end from Lai et al. (2007, 2008).  We tabulate these most-likely results in parenthesis in \tab{singlefit}.

Although we have tried to ensure that our stellar population fits are valid, there are still some concerns with a few objects.  Objects CS2-3, CS2-4 and CS2-6 have {\it Spitzer} fluxes which are blended in with near-neighbors.  While we have used fluxes from TFIT to combat this problem, there is still some uncertainty in the results from these objects.  As we mention in \S 4.4, objects CS2-5, CS2-7 and CS2-8 are undetected in all {\it Spitzer} bands, thus their stellar populations are being derived from the ACS and narrowband fluxes alone (with the NIR and IR upper limits adding some constraints).  To see how these uncertainties affect our conclusions, we look at the results from the other eight objects, which we consider a ``conservative'' sub-sample.  Using the most-likely results from this sub-sample, we find a median q value of 1.00, with q $<$ 1 in 50\% of our objects, close to what we found with the whole sample (71\%).  The median dust extinction in this conservative sub-sample is 2.0 mag, which is less than the sample as a whole.  We still find dust in all objects, but many of those objects with the highest extinction levels have been cut from this sample, thus it is possible that those high values of \dust~may not be accurate.  However, the presence of dust in all objects still implies that these objects are not primitive.  Figure 9 shows the cumulative age distribution for this sub-sample only.  More models fall in the young regime with this sub-sample, as CS2-4 is not included.  This indicates that we do need a larger sample to learn just how significant (if at all) this bimodality is.

\subsection{Causes of Model Uncertainties}
\subsubsection{Importance of NIR}
We spent a lot of effort to try to get meaningful NIR fluxes for our objects.  Although the ISAAC observations ended up not being deep enough, we were still able to put upper-limit constraints in this regime, which did somewhat constrain the models.  However, our results would be much improved with NIR detections.   The reason for this is that the best indicator of age is the spectral break near 4000 \AA.  Objects dominated by older stellar populations will show a significant break here, while young objects will not.  At z $\sim$ 4.5, the NIR brackets this break, with the J and H-band blueward, and the Ks-band just redward.  The wavelength distance between Ks and 3.6$\mu$m is too large for the IRAC band to constrain the break, and it is further complicated by $H\alpha$ emission at this redshift.  Likewise, if dust is involved, the z'-band is too far away to constrain the continuum on the blue-side.  Thus NIR is crucial to constraining the ages of these objects.

CS2-4 has an extremely red z' - 4.5 $\mu$m color (3.24 mag).  With the absence of intervening information, this color is equally likely to be fit by either a lot of dust, or old stars creating a break.  The best-fit model happened to fit it with young stars with \dust~= 5.0 mag.  However, as shown by the Monte Carlo results, a much older age appears more likely, with the break causing the red color.  If one is going to spend future observing time constraining the physical properties of \lya~galaxies, this then shows that the NIR is the most crucial place to observe.

\subsubsection{$H\alpha$ Emission}
As we discussed in F08a, the amount of $H\alpha$ emission relative to the amount of \lya~emission we observe can be a crucial constraint on the cause of the \lya~EW.  If the \lya~EW is high due to young stars, the $H\alpha$~EW will also be high, as both types of photons are created in large numbers in star forming regions.  However, if the \lya~EW is high due to clumpy dust, the $H\alpha$~EW will not be high, as $H\alpha$ photons are not resonantly scattered.  Thus, a high \lya~to~$H\alpha$ ratio implies clumpy dust enhancement of the \lya~EW.

We can gain a handle on this by examining the 3.6 - 4.5 $\mu$m colors.  \fig{halpha}~plots the narrowband excess (F606W - NB) vs. the 3.6 - 4.5 $\mu$m color, which is analogous to $H\alpha$~EW.  We can use this plane to examine whether just looking at the two EWs gives any real physical insight.  To do this, we plot both the positions of our objects in this plane, as well as a few representative model tracks.  The extent of these model tracks represents the amount of dust extinction, from 0 - 5 mag.  The line-style represents the age, with the solid line being a 5 Myr old population, dotted 20 Myr, and dashed 800 Myr.  The line colors represent the clumpiness, with blue being q = 0, green q = 1, and red q = 10.  We then consider the best-fit age, \dust, and q for each object, and see whether we find any trends that are consistent with the models.

Looking at age first, we see that the tracks move to the right with age, representing $H\alpha$ emission being reduced and the optical continuum becoming redder at older ages.  We should thus see our younger objects towards the left, and older towards the right.  However, we do not see this trend, as the oldest three objects, CHa-3, CS2-4 and CS2-6, lie across the whole range of 3.6 - 4.5 $\mu$m color, thus it does not appear as if these two colors alone are enough to constrain the age in practice.

As for the clumpiness, the models show that low q results in a higher narrow-band excess, and vice versa, as we would intuitively expect.  Looking at our objects, we do see this trend, as our three objects with the highest q values have the lowest narrowband excess, and nearly all of the q = 0 objects lie towards the top of this plane.  Lastly, we look at the dust extinction.  If q is near zero, the points should move upward with dust.  If q is near 1, they should move to the right, and if q is very high, they should move directly down.  We do see some trends, although not as strong as those seen with q.  If we take all points with q = 0, we do see a trend of less dust with decreasing narrowband excess, with the exception of CS2-6.  Only two objects have q = 0.75, but we see the same trend with these two.

The increasing trend towards lower redshift science opens up the door to use the $H\alpha$ line as a diagnostic tool to understand LAEs.  We have shown that even at high redshift, the derived \lya-to-$H\alpha$ ratio can put some constraints on the dust extinction and geometry in LAEs.  Future studies with a broader wavelength coverage and more data points (i.e. including the NIR) will allow for better fits to the data.  This will remove some of the uncertainty coming from the model fitting, perhaps allowing the EW comparison to constrain the age.

\subsubsection{Other Scenarios for Preferential \lya~Escape}

While we have shown that clumpy dust can enhance the \lya~EWs in some of our objects, other mechanisms might also explain their observed SEDs.  First, \lya~could appear brighter relative to the continuum if an external gas shell around the star formation region preferentially allowed the escape of \lya\ in some directions.  This would result in \lya~being ``beamed'' in some direction, and if we were to observe the galaxy from this direction, we would detect a large \lya~EW.  This mechanism could result in the \lya~EW being enhanced by a factor of a few, in a modest fraction of galaxies.

A second possibility for \lya~escape involves the presence of large outflows.  These would work in favor of \lya~escape in two ways.  First, if the \lya~photons were to back-scatter off the far side of the expanding shell, they would appear redshifted to the static part of the galaxy during their trek back though the galaxy (i.e. moving mirror effect), and thus would not be scattered by any HI atoms.  Likewise, imagine if the \lya~photons were emitted in our direction from some stationary point in the galaxy, but the near side of the galaxy had an expanding outflow, the \lya~photon would again appear redshifted to the hydrogen in the ISM, and thus would pass out of the galaxy unobscured.  Velocity shifts in the \lya~line relative to interstellar lines have been seen locally by Kunth et al. (1998).  Using a stack of 811 z $\sim$ 3 Lyman break galaxy spectra, Shapley et al. (2003) detected a mean \lya~redshift of 360 km s$^{-1}$ relative to the systemic redshift of the galaxy.  However, \lya~galaxies are intrinsically fainter, and thus it is difficult to detect their continua in individual spectra (i.e., Dawson et al. 2004), let alone interstellar absorption lines.  This remains a top priority for the next generation of extremely large telescopes.

\section{Conclusions}
We have presented the results from our analysis of 10 newly discovered, and four previously known, narrowband selected \lya~galaxies in the CDF--S.  We compared the SEDs of these objects to stellar population models in order to determine their physical properties such as age, mass, and dust extinction.  More specifically, we are interested in finding out whether enhancement of the \lya~EW due to a clumpy, dusty ISM can be responsible for some of the large \lya~EWs which we have observed.

We first computed the best-fit stellar population model to each object allowing one SFH.  For the four objects which we previously analyzed in F08a, we confirmed our earlier results, which had three of the objects being fairly young, and one being very old.  The old object (CHa-3) looks to still have a strong \lya~line due to dust enhancement, shown by its value of the clumpiness parameter (q) of zero.  Although none of the rest of our sample was best-fit by an age over 50 Myr, the majority of our objects appeared to require some amount of clumpy dust enhancement of the \lya~EW, with 9/14 objects having q $<$ 1, thus dust enhancement is widespread.

As a test, we also allowed objects to be fit by two bursts of star formation: One at a maximally old age of 1.4 Gyr, and one at any time.  Most of our objects showed significantly worse fits with two bursts.  However, we do feel as if it was a necessary exercise to fit all objects to this type of model, as a two-burst population could be a viable alternative explanation to those objects which appear old due to dust enhancement.

\begin{figure}
\epsscale{1.2}
\plottwo{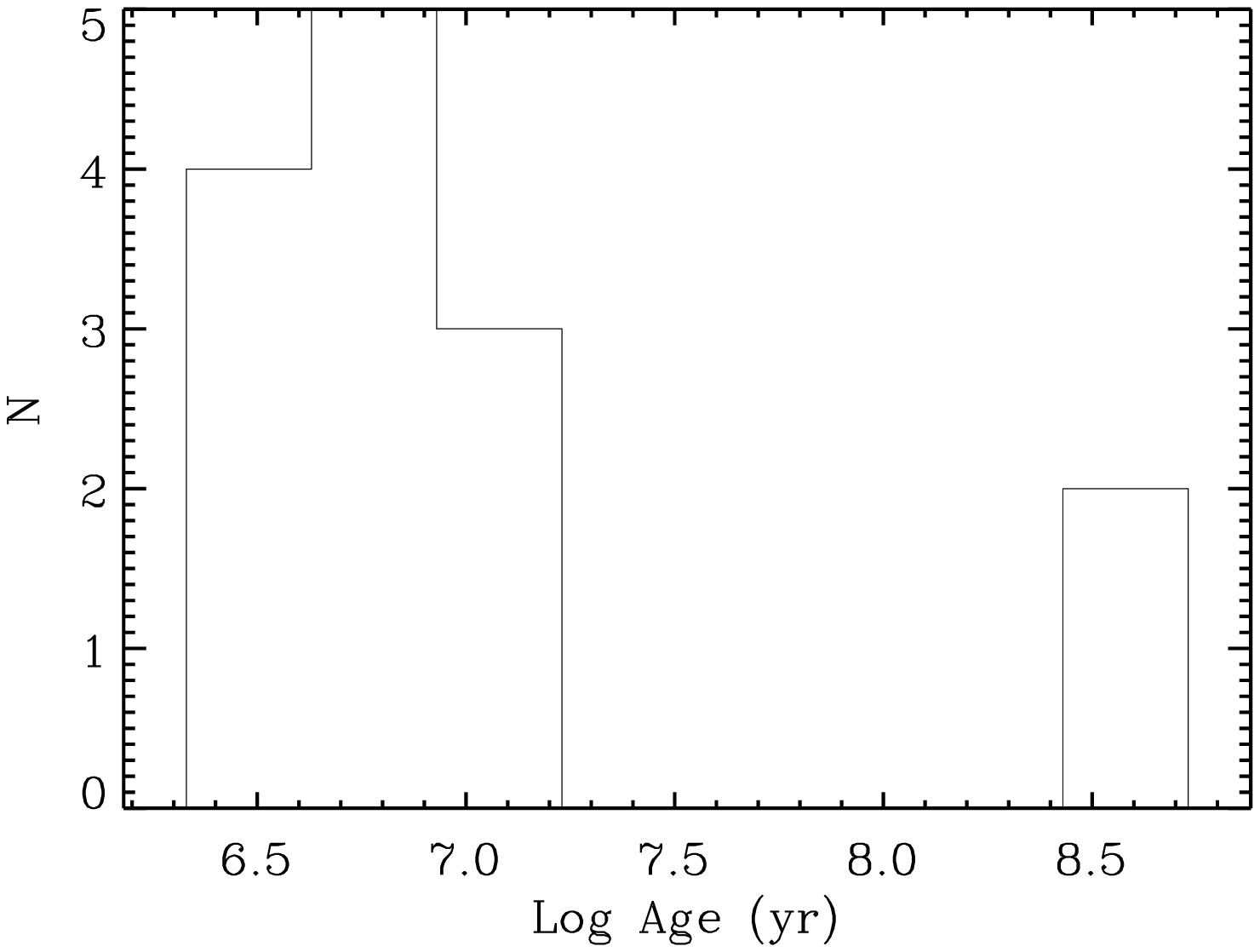}{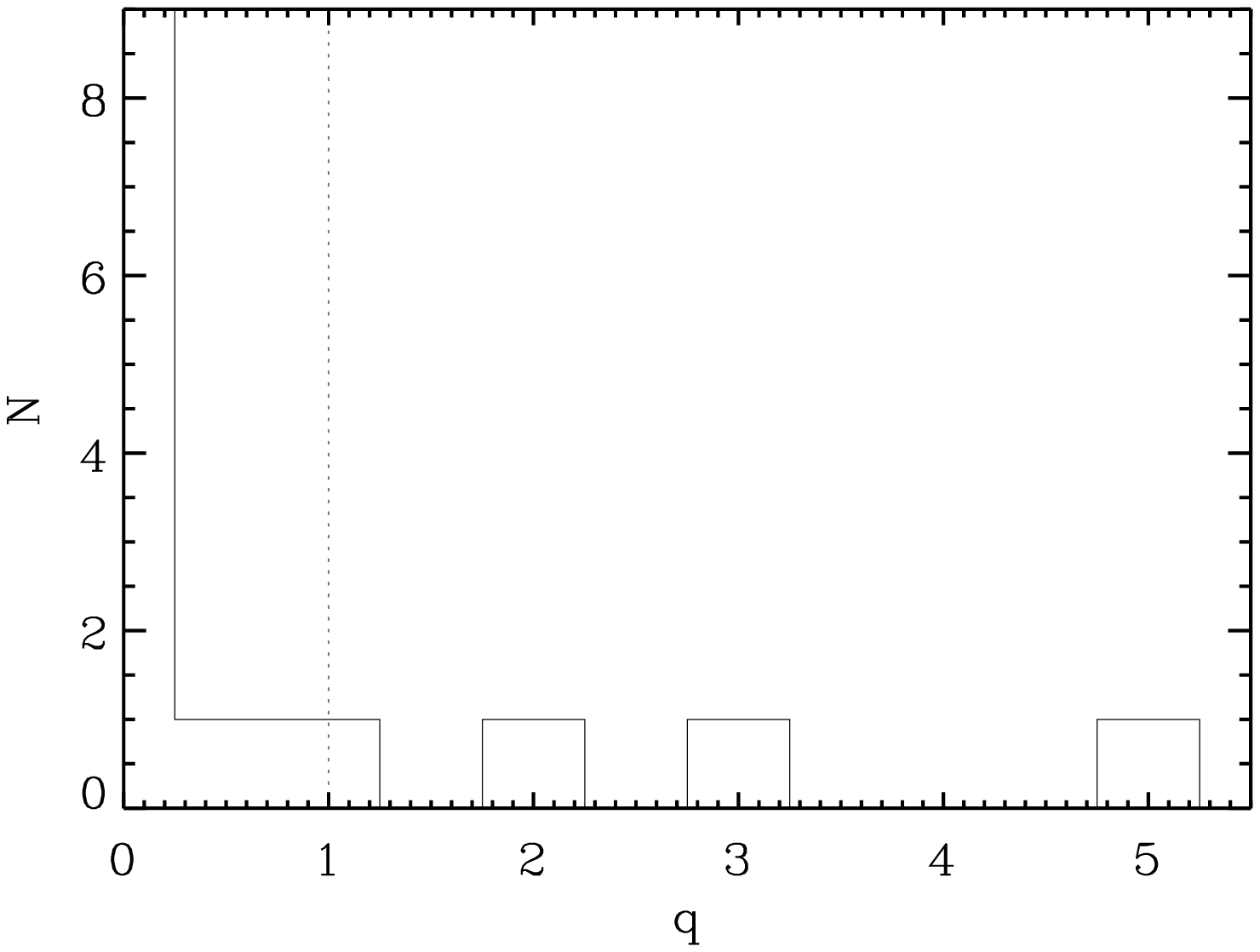}
\caption{The distribution of most-likely stellar population ages from our single component models.  These were derived from the Monte Carlo simulation results.  In cases where the best-fit model did not lie in the largest 68\% confidence region, we set the best-fit age and q value to that derived from the center of the largest 68\% confidence region, computing these new histograms.  The age distribution is interesting, as it implies that some mechanism is preventing us from seeing LAEs at ``moderate'' ages.}\label{fithist2}
\end{figure}

\begin{figure}
\epsscale{1.0}
\begin{center}\plotone{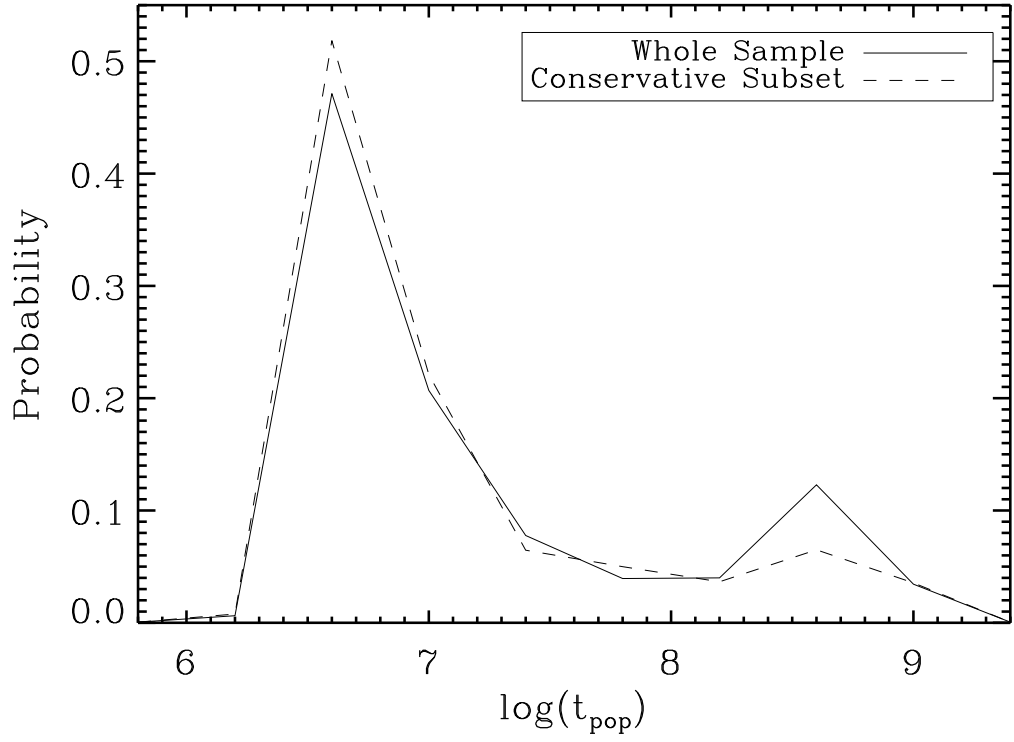}\end{center}
\caption{The composite age probability curve of all of our LAEs, made by averaging the probability curve for each object derived from the Monte Carlo simulation results.  This age probability distribution backs up our by-eye results from \fig{fithist2}a, showing a double peaked distribution, at $\sim$ 4 and 400 Myr.  The dashed line shows the same curve, but only for the eight LAEs in the conservative subsample.}\label{prob}
\end{figure}

\begin{figure}
\epsscale{1.0}
\begin{center}\plotone{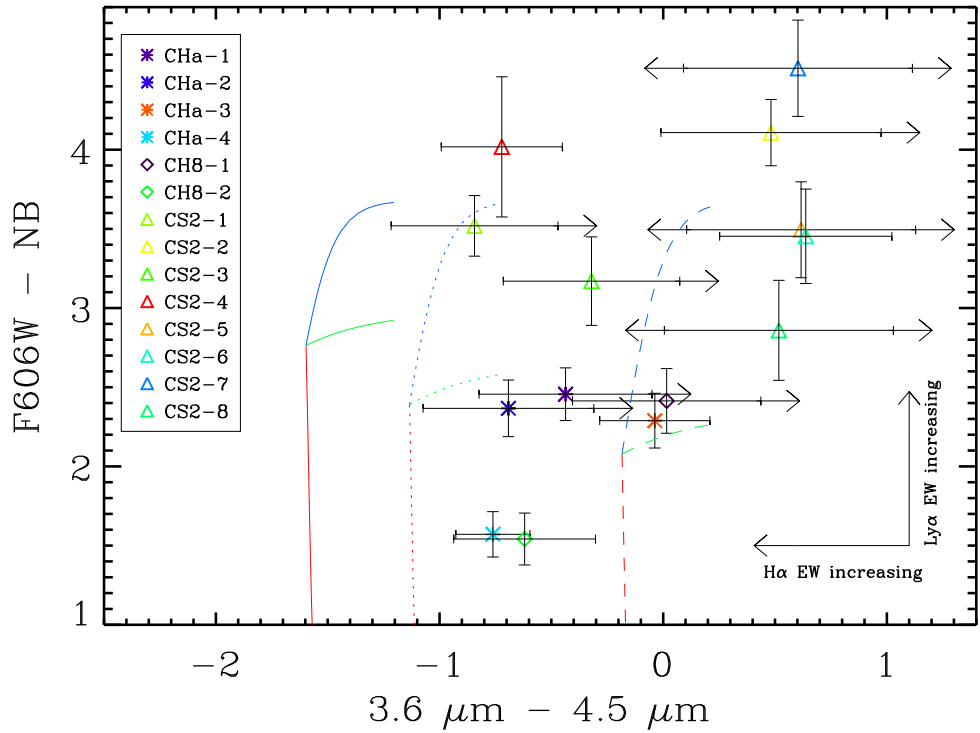}\end{center}
\caption{Color-color plot of our objects, with the narrow-band excess (F606W - NB) on the vertical axis, and the 3.6 - 4.5 $\mu$m color on the horizontal axis, which is a measure of the $H\alpha$ line strength.  Each of our objects are plotted with a different color-symbol combination for identification.  The colors follow the age, with purple being the youngest, and red being the oldest (using the most-likely ages).  We plot arrows showing the direction in the plane that both the \lya~and $H\alpha$~EWs increase.  The colored lines show how our models change with age, dust and q.  All models are 0.02$Z$\sol~and have a continuous SFH.  The colors represent different q values, with blue representing q = 0, green q = 1 and red q = 10.  The line-styles represent different ages, with solid being 5 Myr, dotted 20 Myr and dashed 800 Myr.  The extent of the lines represent the amount of dust extinction, with the crossing of all three colors representing zero dust, and the end of the line representing \dust~= 5.0 mag.  While we do see best-fit q and \dust~trends matching up with the model tracks, these colors do not appear to well-constrain the age, although this may be due to the uncertainty in our model results.}\label{halpha}
\end{figure}

To assess the validity of our results, we ran 7000 Monte Carlo simulations, obtaining a best-fit for each object from each simulation.  In the resulting contour plots, we can examine whether the best-fit model truly represents the most likely model by seeing if the best-fit lies in the largest 68\% confidence region.  If it does not, we assign the object a new age, \dust~and q based on the largest 68\% confidence region.  \fig{fithist2}a and \fig{fithist2}b plot histograms of the most likely ages and q's for our sample.  The distribution of ages is interesting, in that it implies that LAEs are either very young ( $<$ 15 Myr), or very old ($>$ 450 Myr).  However, as Figure 9 shows, we need a much larger sample before we can see if the bimodality is significant.

There could be many explanations for this, but we propose that this bimodality in LAE stellar population ages may be due to dust.  At the beginning of this work, we asked the question of whether LAEs were primitive, dusty or evolved galaxies.  This work, among others, has shown that while many LAEs are young, they are not primitive.  Using our most likely results, we find a range of \dust~of 0.3 to 4.5 mag, thus all of our objects have some amount of dust extinction in them.  As dust comes primarily from evolved stars and stellar deaths, the existence of dust is strong evidence that LAEs are not primitive, as the dust has been produced by a previous generation of stars.  We find that out of our 14 candidate LAEs, 12 of them appear to be dusty star-forming galaxies, with ages from 3 - 15 Myr, and \dust~from 0.4 to 4.5 mag.  The remaining two objects appear to be evolved galaxies, with ages of 450 and 500 Myr, and \dust~of 0.3 and 3.5 mag respectively, exhibiting \lya~emission due to dust enhancement of the EW.

The young galaxies in our sample, although they are dusty due to a previous generation of stars, still manage to emit \lya, mainly due to dust enhancement (8/12 have q $<$ 1).  After a few 10's of Myr, enough massive stars have exploded to further saturate the ISM with dust, and this could explain the drop-off in numbers (\fig{fithist2}a), and why we don't see any LAEs from 15 - 450 Myr.  After some period of time, the stars have changed the ISM geometry enough so that \lya~can escape again.  This ISM is now very patchy, which is why both of our old LAEs have dust enhancement of the \lya~EW in their model spectra.  While this scenario is intriguing, a much larger sample is needed before we can see if this age bimodality is statistically significant.

While we inferred many properties about LAEs in this work, we have also learned that more data is needed before we can truly match our objects to the models.  For most of our objects, we are missing data in a crucial area of the SED, constraining the 4000 \AA~break.  Without these data points, objects are allowed to be either old or dusty to explain the red colors, a degeneracy which can be fixed with better NIR data.  Future observatories such as the {\it James Webb Space Telescope} will be sensitive in this regime, and will provide the data needed to better constrain these objects.  Nonetheless, with the data in hand, we can now say that dust enhancement of the \lya~EW appears to be occurring in the majority of our LAEs, and this effect should be considered in future stellar population studies.

\acknowledgements 
Support for this work was provided in part by NASA through grant numbers HST-AR-11249, HST-GO-10240 and HST-GO-10530 from the SPACE TELESCOPE SCIENCE INSTITUTE, which is operated by the Association of Universities for Research in Astronomy, Inc., under NASA contract NAS5-26555. This work was also supported by the Arizona State University (ASU) Department of Physics and the ASU School of Earth and Space Exploration. We thank Keely Snider, Russell Ryan and Seth Cohen for helpful conversations which greatly improved this paper.

\clearpage
\begin{deluxetable}{cccccccccc}
\tabletypesize{\tiny}
\tablecaption{Optical Magnitudes of Ly$\alpha$ Galaxy Candidates}
\tablewidth{0pt}
\tablehead{
\colhead{Name} & \colhead{GOODS} & \colhead{RA} & \colhead{Dec} & \colhead{Narrow-} & \colhead{F435W} & \colhead{F606W} & \colhead{F775W} & \colhead{F850LP} & \colhead{Rest EW}\\
\colhead{$ $} & \colhead{v1.9 ID} & \colhead{(J2000)} & \colhead{(J2000)} & \colhead{band} & \colhead{$B$} & \colhead{$V$} & \colhead{$i'$} & \colhead{$z'$} & \colhead{(\AA)}\\
}
\startdata
CHa-1&7210&53.066629&-27.708769&24.03 $\pm$ 0.09&$>$ 29.10&26.49 $\pm$ 0.14&25.70 $\pm$ 0.10&25.95 $\pm$ 0.17&189.7$^{+84.1}_{-52.5}$\\
CHa-2&24722&53.165712&-27.854153&24.15 $\pm$ 0.11&$>$ 29.10&26.52 $\pm$ 0.14&25.52 $\pm$ 0.09&25.60 $\pm$ 0.10&166.5$^{+77.9}_{-48.9}$\\
CHa-3&33166&53.243251&-27.894329&24.08 $\pm$ 0.11&$>$ 29.10&26.37 $\pm$ 0.13&25.54 $\pm$ 0.09&25.70 $\pm$ 0.11&149.1$^{+64.6}_{-42.4}$\\
CHa-4&29436&53.201016&-27.860250&24.24 $\pm$ 0.12&$>$ 29.10&25.81 $\pm$ 0.09&24.96 $\pm$ 0.09&24.82 $\pm$ 0.08&56.3$^{+17.2}_{-13.8}$\\
CH8-1&29775&53.204198&-27.817216&24.44 $\pm$ 0.15&$>$ 29.10&26.85 $\pm$ 0.14&26.45 $\pm$ 0.16&26.15 $\pm$ 0.14&175.9$^{+98.3}_{-59.2}$\\
CH8-2&31908&53.225148&-27.833556&24.39 $\pm$ 0.16&$>$ 29.10&25.93 $\pm$ 0.04&24.95 $\pm$ 0.03&25.07 $\pm$ 0.03&53.4$^{+15.7}_{-13.7}$\\
CS2-1&3533&53.039731&-27.773229&23.58 $\pm$ 0.13&$>$ 29.10&27.10 $\pm$ 0.14&25.76 $\pm$ 0.07&25.79 $\pm$ 0.08&2798.8$^{+\infty}_{-1935.2}$\\
CS2-2&7326&53.067369&-27.812334&23.49 $\pm$ 0.14&$>$ 29.10&27.60 $\pm$ 0.16&26.79 $\pm$ 0.13&26.53 $\pm$ 0.11&$>$10000.0$^{+\infty}_{-13279.3}$\\
CS2-3&13009&53.102239&-27.793217&24.12 $\pm$ 0.20&$>$ 29.10&27.29 $\pm$ 0.19&26.63 $\pm$ 0.18&26.01 $\pm$ 0.12&690.9$^{+3648.2}_{-383.5}$\\
CS2-4&16270&53.119230&-27.932951&24.66 $\pm$ 0.25&$>$ 29.10&28.68 $\pm$ 0.36&27.99 $\pm$ 0.30&27.83 $\pm$ 0.29&$>$10000.0$^{+\infty}_{-8424.9}$\\
CS2-5&19333&53.135419&-27.729731&24.68 $\pm$ 0.19&$>$ 29.10&28.17 $\pm$ 0.24&26.84 $\pm$ 0.12&27.31 $\pm$ 0.21&2366.7$^{+\infty}_{-1810.0}$\\
CS2-6&19934&53.138953&-27.695547&24.40 $\pm$ 0.18&$>$ 29.10&27.85 $\pm$ 0.24&27.73 $\pm$ 0.33&27.39 $\pm$ 0.29&1870.8$^{+\infty}_{-1354.9}$\\
CS2-7&24214&53.162613&-27.803605&24.20 $\pm$ 0.19&$>$ 29.10&28.71 $\pm$ 0.24&28.13 $\pm$ 0.24&28.06 $\pm$ 0.25&$>$10000.0$^{+\infty}_{-11292.0}$\\
CS2-8&25038&53.167696&-27.886145&24.96 $\pm$ 0.24&$>$ 29.10&27.82 $\pm$ 0.21&27.09 $\pm$ 0.17&27.20 $\pm$ 0.22&356.3$^{+676.9}_{-183.6}$\\
\enddata
\tablecomments{Optical magnitudes of our LAE candidates.  The narrowband magnitudes come from our ground-based data, while the optical magnitudes come from the GOODS v1.9 catalog (all magnitudes are computed from the SExtractor MAG$\_$AUTO parameter, and thus are a close approximation of the total flux).  The calculated rest-frame EWs assume a redshift equal to that of \lya~at the center of their respective filters.  In three of the objects, the computed EW comes out to be $>$ 10000 \AA~due to the extremely faint continuum (F606W) fluxes.  Likewise, these faint continuum fluxes can result in infinite upper limits on these EWs.  Both CS2-2 and CS2-7 have EWs technically consistent with zero, but these errors are dominated by the faint continuum fluxes.}\label{optmag}

\end{deluxetable}

\begin{deluxetable}{ccccccccc}
\tabletypesize{\small}
\tablecaption{NIR and IR Magnitudes of Ly$\alpha$ Galaxy Candidates}
\tablewidth{0pt}
\tablehead{
\colhead{Name} & \colhead{J} & \colhead{H} & \colhead{Ks} & \colhead{3.6$\mu$m} & \colhead{4.5$\mu$m} & \colhead{5.8$\mu$m} & \colhead{8.0$\mu$m} & \colhead{Redshift}\\
\colhead{$ $} & \colhead{$ $} & \colhead{$ $} & \colhead{$ $} & \colhead{$ $} & \colhead{$ $} & \colhead{$ $} & \colhead{$ $} & \colhead{Information}\\
}
\startdata
CHa-1&$>$ 25.75$^{*}$&$>$ 25.25$^{*}$&$>$ 25.25$^{*}$&25.26 $\pm$ 0.14&$>$ 25.69$^{*}$&$>$ 23.69$^{*}$&$>$ 23.53$^{*}$&4.24\\
CHa-2&$>$ 25.75$^{*}$&$>$ 25.25$^{*}$&$>$ 25.25$^{*}$&25.29 $\pm$ 0.12&$>$ 25.98$^{*}$&$>$ 23.88$^{*}$&$>$ 23.74$^{*}$&4.44\\
CHa-3&$>$ 25.75$^{*}$&$>$ 25.25$^{*}$&$>$ 25.25$^{*}$&24.94 $\pm$ 0.13&24.98 $\pm$ 0.21&$>$ 23.42$^{*}$&$>$ 23.28$^{*}$&4.42\\
CHa-4&$>$ 25.75$^{*}$&$>$ 25.25$^{*}$&$>$ 25.25$^{*}$&24.22 $\pm$ 0.05&24.98 $\pm$ 0.16&$>$ 23.76$^{*}$&$>$ 23.65$^{*}$&4.36\\
CH8-1&$>$ 25.75$^{*}$&$>$ 25.25$^{*}$&$>$ 25.25$^{*}$&25.61 $\pm$ 0.22&$>$ 25.59$^{*}$&$>$ 23.60$^{*}$&$>$ 23.39$^{*}$&3.82\\
CH8-2&$>$ 25.75$^{*}$&$>$ 25.25$^{*}$&$>$ 25.25$^{*}$&24.99 $\pm$ 0.11&25.60 $\pm$ 0.30&$>$ 23.77$^{*}$&$>$ 23.65$^{*}$&-1.00\\
CS2-1&---&---&---&24.87 $\pm$ 0.09&$>$ 25.71$^{*}$&$>$ 23.72$^{*}$&$>$ 23.57$^{*}$&-1.00\\
CS2-2&$>$ 25.75$^{*}$&$>$ 25.25$^{*}$&$>$ 25.25$^{*}$&26.14 $\pm$ 0.33&$>$ 25.66$^{*}$&$>$ 23.56$^{*}$&$>$ 23.37$^{*}$&4.52\\
CS2-3&$>$ 25.75$^{*}$&$>$ 25.25$^{*}$&$>$ 25.25$^{*}$&25.32 $\pm$ 0.16&$>$ 25.64$^{*}$&$>$ 23.58$^{*}$&$>$ 23.42$^{*}$&4.25\\
CS2-4&$>$ 25.75$^{*}$&$>$ 25.25$^{*}$&$>$ 25.25$^{*}$&23.87 $\pm$ 0.08&24.59 $\pm$ 0.26&$>$ 22.84$^{*}$&$>$ 22.74$^{*}$&-1.00\\
CS2-5&$>$ 25.75$^{*}$&$>$ 25.25$^{*}$&$>$ 25.25$^{*}$&$>$ 25.94$^{*}$&$>$ 25.32$^{*}$&$>$ 23.28$^{*}$&$>$ 23.15$^{*}$&-1.00\\
CS2-6&$>$ 25.75$^{*}$&$>$ 25.25$^{*}$&$>$ 25.25$^{*}$&25.08 $\pm$ 0.27&24.44 $\pm$ 0.27&$>$ 22.71$^{*}$&$>$ 22.56$^{*}$&-1.00\\
CS2-7&$>$ 25.75$^{*}$&$>$ 25.25$^{*}$&$>$ 25.25$^{*}$&$>$ 26.48$^{*}$&$>$ 25.88$^{*}$&$>$ 23.87$^{*}$&$>$ 23.72$^{*}$&-1.00\\
CS2-8&$>$ 25.75$^{*}$&$>$ 25.25$^{*}$&$>$ 25.25$^{*}$&$>$ 25.78$^{*}$&$>$ 25.26$^{*}$&$>$ 23.18$^{*}$&$>$ 23.04$^{*}$&-1.00\\
\enddata
\tablecomments{$^{*}$ 3 $\sigma$ upper limits.  Magnitudes of our candidates in the near-infrared (NIR) and IR.  All of our candidates were undetected in the VLT/ISAAC NIR data, thus we display the 3 $\sigma$ upper limits (CS2-1 was not covered by VLT/ISAAC).  When the {\it Spitzer}/IRAC observations had a less than 3 $\sigma$ significance, we display the 3 $\sigma$ upper limits.  The redshift information comes from the MUSIC and FIREWORKS catalogs; see \S 2.4 for more details (-1.00 means that the object was not detected by either survey).}\label{irmag}

\end{deluxetable}
\clearpage
\begin{deluxetable}{ccccccccc}
\tabletypesize{\small}
\tablecaption{Best-Fit Single Population Model}
\tablewidth{0pt}
\tablehead{
\colhead{Name} & \colhead{t$_{pop}$} & \colhead{Mass} & \colhead{Z} & \colhead{$\tau_{SFH}$} & \colhead{A$_{1200}$} & \colhead{q} & \colhead{Model EW} & \colhead{$\chi^{2}_{r}$}\\
\colhead{$ $} & \colhead{(Myr)} & \colhead{(10$^{7}$ M\sol)} & \colhead{(Z\sol)} & \colhead{(yr)} & \colhead{(mag)} & \colhead{$ $} & \colhead{(\AA)} & \colhead{$ $}\\
}
\startdata
CHa-1&3.0&28.12&0.2&4x10$^{9}$&1.00&2.00&81.49&0.52\\
CHa-2&50.0 (4.0)&135.53 (52.84)&0.02&10$^{8}$&1.25 (2.00)&1.00 (1.00)&75.54&0.51\\
CHa-3&453.5&383.94&0.2&4x10$^{9}$&0.30&0.00&97.43&0.32\\
CHa-4&6.0&58.28&0.005&10$^{7}$&1.00&3.00&27.39&1.94\\
CH8-1&2.5 (12.0)&29.16 (157.11)&1.0&10$^{7}$&2.00 (4.00)&1.00 (0.25)&131.32&0.93\\
CH8-2&10.0&45.72&0.005&10$^{7}$&0.50&5.00&20.04&1.07\\
CS2-1&13.2&128.93&0.02&10$^{7}$&2.50&0.75&135.83&9.60\\
CS2-2&13.2&205.40&0.005&10$^{6}$&3.50&0.00&398.16&2.23\\
CS2-3&12.0&605.09&0.005&10$^{5}$&4.50&0.25&340.76&0.74\\
CS2-4&4.0 (500.0)&100.86 (4972.0)&0.005&10$^{6}$&5.00 (3.50)&0.75 (0.00)&478.33&0.82\\
CS2-5&15.1&16.74&1.0&10$^{7}$&1.00&0.00&122.85&3.28\\
CS2-6&40.0 (7.0)&476.58 (94.79)&1.0&10$^{7}$&4.00 (4.50)&0.00 (0.00)&999.58&0.07\\
CS2-7&5.0&16.08&1.0&10$^{5}$&3.00&0.00&533.27&1.47\\
CS2-8&13.2 (7.0)&8.39 (32.85)&1.0&4x10$^{9}$&0.80 (3.00)&0.00 (0.00)&140.79&0.01\\
\enddata
\tablecomments{The minimized $\chi^{2}$ single-population model for each object.  Age, mass, metallicity, SFH timescale, A$_{1200}$ and q (clumpiness) were allowed to vary during the fitting process.  The model EWs are rest-frame (including the effects of IGM absorption), and computed from a ratio of \lya~flux to continuum flux redward of the line.  Values in parenthesis show the most-likely model results (see \S 5.2.2).}\label{singlefit}

\end{deluxetable}
\begin{deluxetable}{ccccccccc}
\tabletypesize{\small}
\tablecaption{Best-Fit Two-Burst Model}
\tablewidth{0pt}
\tablehead{
\colhead{Name} & \colhead{t$_{pop}$} & \colhead{Mass} & \colhead{Z} & \colhead{A$_{1200}$} & \colhead{q} & \colhead{Mass Fraction} & \colhead{Model EW} & \colhead{$\chi^{2}_{r}$}\\
\colhead{$ $} & \colhead{(Myr)} & \colhead{(10$^{7}$ M\sol)} & \colhead{(Z\sol)} & \colhead{(mag)} & \colhead{$ $} & \colhead{in Old Stars} & \colhead{(\AA)} & \colhead{$ $}\\
}
\startdata
CHa-1&2.5&106.34&0.02&1.00&2.00&0.70&72.38&0.53\\
CHa-2&6.0&304.82&0.005&1.75&1.00&0.80&66.99&0.51\\
CHa-3&4.0&1322.55&0.4&0.40&1.00&0.99&66.12&0.40\\
CHa-4&3.0&570.48&0.2&1.50&2.00&0.90&27.77&3.22\\
CH8-1&4.0&86.15&0.2&1.50&0.75&0.80&138.21&0.94\\
CH8-2&4.0&226.99&0.02&0.40&5.00&0.90&21.04&0.96\\
CS2-1&6.9&423.15&0.02&3.00&0.50&0.60&154.35&10.72\\
CS2-2&12.0&282.98&0.005&3.50&0.00&0.30&382.27&2.25\\
CS2-3&12.0&651.26&0.005&4.50&0.25&0.10&340.77&0.74\\
CS2-4&3.0&5109.95&0.005&4.50&0.75&0.99&431.89&1.03\\
CS2-5&3.0&5.24&1.0&0.90&0.50&0.00&108.41&3.38\\
CS2-6&9.1&509.04&0.2&4.50&0.00&0.70&638.01&0.16\\
CS2-7&5.0&15.39&1.0&3.00&0.00&0.00&533.27&1.47\\
CS2-8&3.0&4.48&1.0&0.90&0.25&0.00&132.75&0.01\\
\enddata
\tablecomments{Same as Table 3, only these models had two bursts of star formation; one maximally old burst at 1440 Myr ago, and a second burst at t$_{pop}$.  We list the mass fraction in the old burst for each best-fit model.}\label{tbfit}

\end{deluxetable}

\clearpage
\clearpage
\clearpage


\end{document}